\documentstyle[preprint,aps]{revtex}
\tightenlines
\begin{document}
\baselineskip=15pt

\title{Driven Interface Depinning in a Disordered Medium}
\author{Heiko Leschhorn$^1$
\footnote{present address: Theoretische Physik III,
Heinrich-Heine-Universit\"at D\"usseldorf, Uni\-ver\-si\-t\"ats\-str.~1,
D-40225 D\"usseldorf, Germany},
Thomas Nattermann$^1$,
Semjon Stepanow$^2$,
and Lei-Han Tang$^1$
\footnote{present address: Department of Physics,
Imperial College, 180 Queen's Gate, London SW7 2BZ, UK}
}
\address{
$^1$Institut f\"ur Theoretische Physik, Universit\"at zu K\"oln,
Z\"ulpicher Str. 77, D-50937 K\"oln, Germany
}
\address{
$^2$Fachbereich Physik, Martin-Luther Universit\"at Halle,
Friedemann-Bach Platz, D-06108 Halle, Germany
}
\date{\today}

\maketitle

\begin{abstract}
The dynamics of a driven interface in a medium
with random pinning forces is analyzed.
The interface undergoes a depinning transition
where the order parameter is the interface velocity $v$,
which increases as
$v \sim (F-F_c)^\theta$ for driving forces $F$
close to its threshold value $F_c$.
We consider a Langevin-type equation which is expected to be valid
close to the depinning transition of an interface in a
statistically isotropic medium.
By a functional renormalization group scheme the critical
exponents characterizing the depinning transition
are obtained to the first order in $\epsilon=4-D>0$, where $D$ is the
interface dimension.
The main results were published earlier [T. Nattermann et al.,
J. Phys. II France {\bf 2} (1992) 1483].
Here, we present details of the perturbative calculation and
of the derivation of the functional flow equation for the
random-force correlator. The fixed point function of the correlator has a
cusp singularity which is related to a finite value of the threshold
$F_c$, similar to the mean field theory.
We also present extensive numerical simulations and compare them
with our analytical results for the critical exponents.
For $\epsilon =1$ the numerical and analytical results
deviate from each other by only a few percent.
The deviations in lower dimensions $\epsilon = 2,3$ are larger
and suggest that the roughness exponent is somewhat larger
than the value $\zeta = \epsilon / 3$ of an interface in thermal equilibrium.
\end{abstract}

\vspace{1truecm}



\section{Introduction}

The ordering dynamics of an Ising magnet following a quench below the
Curie (or Neel) temperature is controlled, to a great extent,
by the motion of domain walls \cite{NatRuj}.
This is believed to be true also for impure
ferromagnets, where the impurity may either be nonmagnetic
(thus we are speaking of dilution or random-bond disorder)
or act as to differentiate energetically the coexisting phases
(thus we are speaking of random-field disorder).
When the disorder is weak, either in terms of impurity strength or
concentration, its effect on the bulk phases is minor
(at least well above the lower critical dimension),
but the behavior of the interface may be dramatically altered:
the interface becomes rough in order to take advantage of the
inhomogeneous medium, and consequently (free) energy
barriers appear between favorable interface configurations, leading to
pinning phenomena and extremely slow activated dynamics.

What happens if we apply an external field $F$ in favor of
one of the coexisting phases? Thermodynamically one phase
prevails, but this is achievable only if the interface
can get out of its locally favorable configurations.
At zero temperature, the expected behavior for the interface
velocity $v$ as a function of the driving force is shown
in Fig. 1. When the driving force is weak, the interface may
adjust itself locally and settles in a nearby metastable position.
Steady interface motion
sets in only above a threshold value $F_c$ for the driving force.
At finite temperatures, the transition from a pinned to a moving
interface is smeared out by
thermally activated motion. At low temperatures the rounding effect
is however weak
apart from a very small region around the transition.

The $v$-$F$ curve shown in Fig. 1 is actually representative of a number
of other threshold phenomena in condensed systems.  Examples include the
electrical transport of charge-density-waves (CDWs)
\cite{EfeLar,FukLee,Fishermft,NarFiscdw}
and driven flux motion in type-II superconductors \cite{LarOvc}.
A basic theoretical question in all these cases is
whether one can calculate the $v$-$F$ curve using a suitable model,
which to a large extent involves an estimate for $F_c$ and a theory for
the singular behavior around the transition.  The first part has been
essentially answered twenty years ago in the pioneering work of
Larkin \cite{Larkin}.
Progress on the second part has been made only recently
\cite{NSTL,NarFisint}.
The difficulty has to do with the fact that the depinning transition
is a nonequilibrium critical phenomenon which involves correlated motion
of all parts of the system.  As it turns out, the key to the solution
lies in properly renormalizing the random-force correlator that
characterizes the medium.

The present paper is an extension of an earlier one where the basic
procedure and results of a functional renormalization group (RG)
treatment of the interface problem were reported \cite{NSTL}.
The strategy we followed is very similar to the ones for treating
equilibrium critical phenomena: the interface model has an
upper critical dimension $D_c=4$ above which perturbation
theory (expanding around a flat, moving interface) converges
at weak disorder. For $D=4-\epsilon<D_c$, it is possible to
transform the perturbation theory into a RG procedure
using momentum shell integration, which yields flow equations
for the parameters of the model. From the fixed point solution of
the flow equations one obtains exponents that characterize the
depinning transition.
Our main aim here is to present details of the perturbative calculation.
We shall also try to clarify the relation between the
interface model used in our analysis and those considered by others.
Finally, we compare the analytical results with those from our
numerical simulations \cite{gimo,PhD}
which indicate some discrepancy in lower dimensions.

The paper is organized as follows. In Sec. II we introduce the model
and review some of its basic properties, in particular the considerations
that lead to the identification
of an upper critical dimension and characteristic lengths.
Section III contains formulation and results of a perturbation
theory. The RG flow equations based on the perturbative calculation
are presented in Sec. IV, along with a discussion of fixed point
solutions and exponents. In Sec. V we present the results of our numerical
study. Conclusions and
some of the open problems are discussed in Sec. VI.

\section{The Model and Some Basic Properties}

\subsection{The model}

On a certain coarse-grained level, the motion of an interface
in an isotropic, weakly disordered medium in the viscous limit can be
described by the following equation for the normal velocity
$v_n({\bf R})$
at a position {\bf R},
\begin{equation}
\label{meqmnor}
\lambda v_n = 2\gamma H+ F + \eta ({\bf R})+\varphi({\bf R},t).
\end{equation}
Here $H$ is the mean curvature of the interface at {\bf R},
$F$ is the driving force (which is actually
a thermodynamic potential that couples to the order parameter),
$\eta({\bf R})$ is a quenched random force,
$\varphi({\bf R},t)$ is a thermal noise, and $\lambda$ and
$\gamma$ are friction (or viscosity) and surface tension coefficients,
respectively.
Such an equation can be motivated from a Ginzburg-Landau
theory for the dynamics of the bulk order parameter
\cite{AllCah,BauDoh,PhD}.
Here we simply accept (\ref{meqmnor})
as a plausible dynamical model for an interface in a disordered
medium that is isotropic in a statistical sense.

We now choose a coordinate system with the $z$-axis perpendicular to the
average interface orientation and the remaining coordinates
forming the $D$-dimensional {\bf x}-space parallel to the interface.
Ignoring overhangs, the interface position is uniquely specified
by a function $z({\bf x},t)$.
Then Eq. (\ref{meqmnor}) becomes
\begin{equation}
\label{meqmsqg}
{\lambda \over \sqrt g} {\partial z \over \partial t} =
\gamma\nabla\cdot (g^{-1/2} \nabla z)+F + \eta ({\bf x} , z)
+ g^{-1/4} \varphi ({\bf x},t),
\end{equation}
where $g = 1 + (\nabla z)^2$.
On sufficiently large scales parallel to the interface,
the factors $g$ on the right-hand-side of (\ref{meqmsqg}) can be set to one.
The effect of the $g$ factor on the left-hand-side of (\ref{meqmsqg})
is more subtle.
When the velocity $v$ of the interface is finite, an expansion
of $g$ in the height gradient gives rise to a term
${1\over 2}\lambda v(\nabla z)^2$
which changes the asymptotic scaling behavior to that
of the Kardar-Parisi-Zhang universality class \cite{KPZ}.
This term vanishes as $v\rightarrow 0$.
Thus we expect an interface model defined by
\cite {Feigelman,BruAep,KopLev}
\begin{equation}
\label{8}
\lambda {\partial z \over \partial t} = \gamma
\nabla ^2 z + F + \eta ({\bf x} , z),
\end{equation}
to have the same critical behavior at the
zero temperature depinning transition as (\ref{meqmsqg}),
provided the gradient of the interface height
tends to zero on large scales.

Equation (\ref{8}) is the final model to be  analyzed in this paper.
At this point a specification of the random force $\eta$
is called for. For random-field disorder, one may
suppose that $\eta({\bf x}, z)$ is a gaussian random variable
that varies over a distance of order $a$ (e.g., the typical
extension of an impurity). A different situation arises
for random-bond disorder, where the random force takes a differential form
$\eta({\bf x},z)=-{\bf n}\cdot\nabla_{\bf R} V({\bf R})
\simeq -\partial V({\bf x},z)/\partial z$, with $V({\bf R})$
being a random potential correlated over a distance $a$
and {\bf n} the interface norm.
A model for driven CDWs can also be casted in the form
(\ref{8}), with $z$ being a phase variable and
$\eta({\bf x},z)=\eta_0\sin [z-\beta({\bf x})]$, where
the independent random variable is
the preferred phase $\beta({\bf x})$ \cite{FukLee,Fishermft,NarFiscdw}.
As far as our analysis goes, the three different situations allow
a unified description in terms of
the pair correlator of the gaussian random forces,
\begin{equation}
\label{mcoreta}
\langle \eta ({\bf x}' , z')
\eta ({\bf x}' + {\bf x} , z'+ z) \rangle =
\Delta_{\parallel} ({\bf x}) ~\Delta (z),
\end{equation}
where, without loss of generality, the mean of $\eta$ is chosen to be zero.
Here and elsewhere $\langle\cdot\rangle$ denotes average over
the disorder distribution.
The correlation of $\eta$ in {\bf x}-space is understood
to extend over a distance of the coarse-graining length $a_{\|}$ and
its precise form is not important in our analysis.
In contrast, the functional form of the random-force correlator
in $z$-space, $\Delta(z)=\Delta(-z)$, turns out to be of crucial importance.

Figure 2 illustrates the appropriate functional forms for $\Delta(z)$
for the three different cases discussed above.
The correlator for the random-field disorder is depicted in
Fig. 2a), where $\Delta (z)$ is a monotonically decreasing function of $z$
for $z>0$ and decays exponentially to zero beyond a distance $a_\perp$.
Figure 2b) shows the correlator for the random-bond disorder,
where $\Delta(z)=-R''(z)$. Here $R(z)$ denotes the correlator of
the random potential $V({\bf x},z)$ in $z$-direction, which
has the same characteristics as the random force correlator
shown in Fig. 2a). For the CDW problem [Fig. 2c)], $\Delta(z)$ is periodic
in $z$ and in addition the integral of $\Delta(z)$ over one
period is zero.

\subsection{The upper critical dimension}

Suppose we accept (\ref{8}) as a reasonable model for the
driven interface, what can one say about its properties
before trying to solve it?
Surely enough, one can arrive at an important conclusion
on the existence of the depinning transition below
a certain critical dimension, and an estimate for the
threshold force $F_c$ in the case of weak disorder, using a simple
argument due originally to Larkin \cite{Larkin} and developed in the
interface context by Bruinsma and Aeppli \cite{BruAep}.
Let us recall the argument below.

We begin with the remark, that a completely flat interface is never pinned.
Indeed, the total random force on a piece of a flat interface of
extension $L$ is roughly $\Delta ^{1/2}(0) L^{D/2}$ as follows
easily from (4). This has to be compared with the total driving force
$FL^D$, which will always win for $L\to \infty$.

For a rough interface this estimate is only valid on scales $L$
at which the typical interface distortion $h$ is smaller than $a_\bot$
and hence pinning is possible for small $F$. However, roughness of
the interface will be suppressed by the curvature force
$\gamma h L^{-2+D}$, which tends to flatten the interface.
In $D>D_c=4$ dimension the curvature force will always win on sufficiently
large length
scales. Thus, for very weak disorder, $\Delta (0)\ll \gamma^2a_{\|}^{D-4}
a^2_{\perp}$, there is no length scale down to $a_{\|}$ at which the
interface is rough and hence there is no pinning $F_c=0$. (Pinning is however
possible for larger disorder).

The situation is different for $D<D_c$: the pinning forces
cannot be balanced by the elastic forces for $h\leq a_{\bot}$
on length scales larger than the Larkin length \cite {Larkin,BruAep},
\begin{equation}
\label{L_c}
L_c = [\gamma^2 a^2_{\bot} / \Delta (0)]^{1/\epsilon},
\end{equation}
where $\epsilon = 4-D$.
Beyond this length, the rough interface explores the random
environment to acquire a finite pinning force per unit area.
The force that is needed to get the interface out of a typical
pinned configuration is thus of the order of \cite{Feigelman,BruAep}
\begin{equation}
\label{F_c}
F_c\simeq \Delta^{1/2}(0)L_c^{-D/2}\sim \Delta^{2/\epsilon}(0).
\end{equation}

\subsection{The depinning transition and scaling}

The argument presented above suggests that,
below the critical dimension $D_c=4$,
there is a depinning transition at a finite strength of
the driving force.
An observation due to Middleton \cite{Middleton}, originally made
in the CDW context, makes it plausible that $v$ is a continuous
function of $F$. Thus the transition, if it exists, should be
continuous and the interface is expected to exhibit critical
fluctuations at the transition and
crossovers away from the transition \cite{NSTL}.

We discussed already, that the interface is rough. In particular,
we will assume, that the interface is self-affine, an assumption which has
to be proven afterwards. In the thermodynamic limit the height correlation
function is assumed to obey the following scaling form
\begin{equation}
\langle [z({\bf x},t)-z({\bf x}',t')]^2\rangle\sim
|{\bf x}-{\bf x}'|^{2\zeta}G\left(\frac{t-t'}{|{\bf x}-{\bf x'}|^z}\right),
\end{equation}
with $G(y) \to $ const for $y \to 0$ and
\begin{equation}
G(y) \sim y^{2\zeta/z}, \ y \to \infty,
\end{equation}
where
$\zeta$ and $z$ (not to be confused with the interface height)
are roughness and dynamic exponents, respectively.

For $F>F_c$ there is a length (and correspondingly, time) scale,
above which one can neglect
the non-linearity of Eq. (3), which is hidden in the random force term.
Going over to a co-moving frame $z=vt+h({\bf x}, t)$, where $v$ is the
mean interface velocity and $\langle h({\bf x},t)\rangle=0$,
we consider first the case
of high velocity. Then $\eta ({\bf x},z)$ can be replaced by
$\eta ({\bf x}, vt)$, i.e. $\eta$ acts as a thermal
noise. In this case the exponents are $z=2$ and $\zeta =0$ for $D>2$
and $\zeta = \frac{2-D}{2}$ for $D<2$ \cite{EW}. Upon decreasing $v$,
the non-linearity becomes more important.
Indeed, if a small part of the interface hits
an obstacle, it will first keep its position, i.e. $h({\bf x},t)$ in the
argument of $\eta({\bf x},vt+h)$ will decrease
by $-\Delta h$ from its original value
to compensate the increase of $vt$.

The typical value of $|\Delta h|$ increases with $t$ according to (7)
and (8) as
$t^{\zeta /z}$. Since $\zeta < 1$ for a well defined interface and $z
\ge 1$
(as we will show below), $h$ changes more slowly than $vt$ and there is
a timescale $t_x \sim v^{-z/(z-\zeta)}$ where the $vt$ term starts to
dominate the argument of $\eta$. On this time scale, the perturbation
of the interface by the obstacle has spread over a region of linear size
\begin{equation}
\xi \sim v^{-\frac{1}{z-\zeta}}.
\end{equation}
For larger time or length scales the effect of the non-linearity can
be neglected and the above mentioned exponents apply.
For smaller scales however, we expect new exponents dominated by the
non-linear terms of (3). If the velocity decreases close to the threshold
as a power law
\begin{equation}
v\simeq (F-F_c)^\theta,
\label{defthe}
\end{equation}
we get $\xi \simeq (F-F_c)^{-\nu}$ with [8]
\begin{equation}
\label{theta}
\theta = \nu (z - \zeta).
\end{equation}

Another important constraint on the exponents follows from
a Harris type argument\cite{Harris} for a homogeneous
transition in a disordered medium:
Indeed,
let us divide the volume which includes the interface
into correlated regions,
each of linear size $\xi$ parallel and $\xi^\zeta$
perpendicular to the
interface, respectively.
The typical fluctuation $\delta F_c$ of the threshold
force $F_c$ in such a region is
\begin{equation}
\delta F_c \sim F_c / \xi ^{(D+\zeta)/2}.
\end{equation}
To have a sharp transition, the condition $\delta F_c \ll F-F_c$ for
$F \to F_c$ has to be fulfilled. This gives, together with (12),
\begin{equation}
\label{harcon}
1/\nu\leq (D+\zeta)/2.
\end{equation}

\subsection{Mean field theory}

Above the upper critical dimension $D_c=4$,
the dimensional analysis of Sec. II.B suggests
absence of a depinning threshold when the disorder is sufficiently weak.
This conclusion can
be verified explicitly using a perturbation theory discussed in the
next section, provided $\eta ({\bf x},z)$ is differentiable in $z$.
An exception is given by the case where
$\eta ({\bf x},z)$ may increase abruptly as $z$ increases
(i.e., the pinning force weakens in a discontinuous way),
leading to a sudden release of an interface segment.
As the phenomenon has an analogy to avalanche motion in
lower dimensions, let us discuss it in some detail below.

For simplicity, let us consider a mean field approximation to Eq. (\ref{8}),
which can be obtained
by taking the limit of infinite-range, suitably normalized elastic
interactions or $D \to \infty$.
Then, all interface elements $ z({\bf x},t) $ are
uniformly coupled to the spatially averaged position $\bar z(t)$
instead of a coupling between nearest neighbors as
in Eq. (\ref{8}).
Assuming that the mean position $\bar z(t)$
moves with a constant velocity $v$, i.e. $\bar z(t) = vt$,
one obtains\cite {Fishermft,KopLev,mft},
\begin{equation}
\label{mft14}
{\partial z \over \partial t} = \tilde \gamma
\bigl [ vt - z(t) \bigr ] + F + \eta (z),
\end{equation}
where we have omitted the redundant $\bf x$-dependence of $\eta$
and set $\lambda = 1$.

Equation (\ref{mft14}) acquires a more transparent form \cite{TanSte}
under a new variable $u \equiv z - vt - (F-v)/\tilde \gamma$
which satisfies (upon a translation of $\eta$),
\begin{equation}
\label{mftnew}
\dot u = -\tilde \gamma u + \eta (vt + u),
\end{equation}
with the self-consistency condition
\begin{equation}
\label{selcon}
\tilde \gamma \langle u \rangle  = -F+v.
\end{equation}
The driving force $F$ now disappears from the equation of motion
(\ref{mftnew}) and can be computed from (\ref{selcon}) as a function of $v$
once a solution to (\ref{mftnew}) is found.

The variable $u(t)$ in
Eq. (\ref{mftnew}) can be interpreted as the position of
an overdamped particle in a moving potential,
\begin{equation}
W(u,t) = {\tilde \gamma \over 2} u^2 + U(vt + u).
\label{potential}
\end{equation}
The random part $U(vt + u) = - \int _{z_0}^{vt+u} dz~\eta(z)$ travels
at a constant velocity $v$ to the left or to the right depending on the sign
of $v$.

Using the overdamped particle language,
one can relate the threshold force to certain
geometrical properties of $\eta$ without an explicit solution
of Eq. (\ref{mftnew}).
Let us define a threshold force for forward motion,
$F_c^{(+)}=\lim_{v\rightarrow 0^+} F(v)$ and
a threshold force for backward motion,
$F_c^{(-)}=\lim_{v\rightarrow 0^-} F(v)$.
For a random force with a statistical inversion symmetry
[i.e. $\eta (z) \to -\eta (-z)$],
$F_c^{(+)} = -F_c^{(-)} \equiv F_c$
and Eq. (\ref{selcon}) yields \cite{TanSte}
\begin{equation}
\label{lim+-}
F_c = {\tilde \gamma \over 2} \Bigl [
\lim _{v \to 0^-} \langle u \rangle -
\lim _{v \to 0^+} \langle u \rangle \Bigr ].
\end{equation}

Fisher\cite{Fishermft} observed that,
in the limit $v \to 0$, the particle stays
in a local minimum of the potential $W$ essentially the whole time,
interrupted by occasional jumps when the minimum it follows disappears.
When the potential travels to the left ($v>0$), the particle
traces the leftmost minimum of $W$ (apart from an initial transient).
The opposite occurs when the potential travels to the right.
Thus $F_c$ is simply proportional to the average distance between the left
and right local minima of the potential $W$.
For a continuous, slow-varying function $\eta(z)$ with a sufficiently
small amplitude, the potential $W$ has a unique minimum all the time,
and hence $F_c=0$.
However, for a discontinuous function $\eta(z)$ (or for sufficiently
strong disorder), the potential $W$ has, from time to time,
more than one minimum, and hence a nonvanishing $F_c$.
Note that only upward jumps in $\eta$ give rise to
upward cusps in the potential which, at times, serve
to break one minimum into two.

The different roles played by upward and downward
discontinuities of $\eta$ when the velocity of the particle is
small can be appreciated by considering
a simple, ``two-state'' model, which has been solved explicitly\cite{mft}.
In this model, the $z$-axis is divided into a uniform
sequence of cells, each of length $a$. The pinning force
$\eta(z)$ is constant in each cell:
$\eta=\pm \eta _0$ with equal probability.
Both upward and downward discontinuities are thus present.
When the particle encounters a downward discontinuity,
it sticks at the cell boundary until the pulling force
[including the $vt$-term in Eq. (\ref{mft14})]
builds up to overtake the resisting force $-\eta_0$ in the cell ahead,
and then enters the new cell at a velocity of the order of $v$.
The situation is different at an upward discontinuity.
When the particle reaches such a point, it suddenly experiences
a pulling force by an amount $2\eta_0$ greater, so that
its velocity jumps by the same amount [see Eq. (\ref{mftnew})].
This relatively fast motion makes it possible for the particle
to sample the negative parts of $\eta$ more often than the positive parts
in the limit $v\rightarrow 0$,
giving rise to a finite $F_c$.
At sufficiently small $\eta_0$
(i.e., no more than
two minima occurs at any given time),
$W$ has two minima
when the upward discontinuity is within a distance $\eta_0/\tilde\gamma$
from the origin.
The distance between the two minima is given by
$\Delta u=2\eta_0/\tilde\gamma$.
Using Eq. (\ref{lim+-}), one finds
$F_c = \eta_0^2/(2 a \tilde \gamma)$.
Apart from the fast motion (i.e. a sort of instability)
caused by the sudden increase of driving at an
upward discontinuity of $\eta$,
the rest of the motion is regular and linear response theory applies.
Hence for $F$ greater than but close to $F_c$ one expects $v\sim F-F_c$.
The mean-field velocity exponent is thus given by $\theta _{\rm MF} = 1$.
The above results for the threshold and for $\theta_{\rm MF}$ are
in agreement with the explicit solution of the model\cite{mft}.

It is interesting to see that
a perturbative treatment of Eq. (\ref{mft14}) \cite{KopLev}
also yields the mean-field exponent $\theta _{\rm MF} = 1$,
although, for the ``two-state'' model, it yields a threshold which is off by
a factor of 2.
In the lowest order perturbation theory, the disorder
is completely specified by its correlator $\Delta(z)$.
The result for the threshold is given by,
\begin{equation}
\label{F_cper}
F_c \simeq -\Delta '(0^+) / \tilde \gamma+O(\Delta^2),
\end{equation}
i.e., for weak disorder,
a finite threshold requires a cusp singularity of the random-force
correlator at distance $0^+$.
For the two-state model,
$ \Delta (z) = \eta_0^2 (1 - |z|/a) $ for $ |z|<a $ and $ \Delta(z) = 0 $.
Both upward and downward discontinuities of $\eta$ contribute to
a nonzero $\Delta'(0^+)=-\eta_0^2/a$.

What can we learn from the above discussion
for the case $D<D_c$, for which
the scaling arguments suggest a finite threshold?
We have seen that for $v \to 0$ the particle $z(t)$ in mean field
theory follows most of the time the minimum of its local potential.
The same is true for an interface element $z({\bf x},t)$ for
$D<D_c$.
The long periods of slow motion of $z({\bf x},t)$ are interrupted
when the local minimum disappears.
This happens
when elastic forces pull the interface element
over a barrier of negative $\eta$'s.
Then the interface element rapidly moves forward and it can itself
pull neighboring interface elements over their barriers (``avalanche'').
In mean field theory
the analogous scenario is when
the particle $z(t)$ is dragged by
the mean position $\bar z (t) = vt$
over an upward discontinuity of the random force.
The following rapid motion of the particle (``jump'')
results in a finite threshold $F_c > 0$.
The jumps and the finite threshold are a consequence of
the  discontinuity in $\eta (z)$, which correspond
to a cusp of the correlator (see Eq. (\ref{F_cper})).
Given the observation that the jumps (and a finite threshold)
{\it are} present for
$D<D_c$, one can expect that this jerky motion
has to be described by
an {\it effective} singular correlator $\Delta (z)$.
In other words, the avalanche-like behavior for $D<D_c$ can be
a mechanism that gives rise to an effective
discontinuous random force acting on the interface.
We will see below in the functional RG treatment,
that the fixed point function of $ \Delta (z)$ has indeed a cusp
singularity at the origin, and that this cusp is responsible
for a finite threshold, as in mean field theory \cite{hand} !

\section{Perturbation theory}

\subsection{Iteration scheme}

In this section we present a perturbation theory which forms the basis
of our renormalization group calculation.

In the moving phase the interface fluctuates around its average
position $vt$, $z({\bf x},t) = vt+h({\bf x},t)$.  Then, Eq. (\ref{8})
can be written as
\begin{equation}
\label{7}
\lambda {\partial h \over \partial t} = \gamma \nabla ^2 h + F -
\lambda v + \eta \bigl({\bf x} , v t + h( {\bf x} , t)\bigl).
\end{equation}
A perturbative solution to Eq. (\ref{7}) can be found by expanding
$\eta({\bf x},z)$ around the flat co-moving reference interface $vt$.
To do so we Fourier-transform $ \eta \bigl({\bf x} , v t + h( {\bf x}
, t)\bigl) $ with respect to its second argument and expand the term $
e^{i q h} $,
\begin{eqnarray}
  \nonumber &&\eta \bigl ( {\bf x},vt+h({\bf x},t)\bigr ) = \int _q
  ~\eta ({\bf x},q) ~e^{iq[vt + h({\bf x},t)]}\\ &&= ~\int _q ~\eta
  ({\bf x},q) ~e^{iqvt} \Bigl( 1 + iqh({\bf x},t) +~{1 \over 2 }
  (iq)^2 [h({\bf x},t)]^2 +...\Bigr),
\label{expeta}
\end{eqnarray}
where $\int _{-\infty} ^\infty dq / (2 \pi) $ is abbreviated by $\int
_q$.  Since the strength of the disorder $\Delta (0)$ determines the
amplitude of the height fluctuations $h$, one expects that the
expansion in powers of $h$ converges for sufficiently weak disorder.

Using (\ref{expeta}), we can rewrite (\ref{7}) in Fourier space,
$$ \tilde h({\bf k}, \omega) = \tilde G_0 ({\bf k}, \omega) \Bigl \{
\delta ({\bf k}) \delta (\omega) ~[F- \lambda v]
+ {1 \over v} \tilde \eta \left({\bf k},{\omega \over v}\right)
+\tilde \varphi({\bf k}, \omega) $$
$$ + \int _q \int _{\bf p} ~i q~ \tilde \eta ({\bf k-p},q) ~\tilde
h({\bf p},\omega - qv) $$
\begin{equation}
  + {1 \over 2} \int _q \int _{\bf p } \int _\Omega \int _{\bf p'}
  ~(iq)^2~
\label{10}
\tilde \eta ({\bf k-p-p'},q) ~\tilde h({\bf p},\Omega) ~\tilde h({\bf
  p'},\omega - \Omega -qv)~~+~~...~~\Bigr \}.
\end{equation}
Here, $\int _{\bf p}\equiv\int d^D p /(2
\pi)^D$
and the bare propagator is defined by
\begin{equation}
\label{10a}
\tilde G_0 ({\bf k}, \omega) = {1 \over \gamma k^2 + i \lambda
  \omega}~.
\end{equation}
We added a source term $\tilde \varphi ({\bf k}, \omega)$ on the
$rhs$ of Eq. (\ref{10}).  The Fourier-transformed quantities, e.g.,
\begin{equation}
  \tilde h ({\bf k}, \omega )= \int d^Dx dt ~h({\bf x},t) e^{-i({\bf
      k\cdot x} + \omega t)}~,
\end{equation}
are marked by a tilde.
Eq. (\ref{10}) can be represented diagrammatically, as shown in Fig.
3. The short double arrow represents $\tilde h$,
a line with an arrow is the bare
propagator $ \tilde G_0({\bf k},\omega) $, a cross $\times$ stand
for $ \delta ({\bf k}) \delta (\omega) ~[F- \lambda v]$ and the
open circle denotes the
source term $\tilde \varphi$.
A filled
circle with a dotted line indicates the quenched noise $ \eta $, where
the number of outgoing arrows of such a vertex
determines the power of the term $ iq $ and the number of $ D+1 $
dimensional integrations.

The averaging over the quenched noise will be represented by the
connection of two dotted lines and is performed using
\begin{equation}
\label{12}
\langle \tilde \eta ({\bf k} , q) \tilde \eta ({\bf k}' , q') \rangle
= \tilde\Delta_\parallel({\bf k}) \Delta _q~\delta ^D ({\bf
  k+k'}) ~\delta (q+q'),
\end{equation}
where $\tilde\Delta_\parallel({\bf k})\simeq e^{-a^2_\parallel{\bf
    k}^2}$ is the Fourier transform of $\Delta_{\parallel}({\bf
  x})$
and
$ \Delta _q $
is defined by
\begin{equation}
  \Delta (z) = \int _q \Delta _q~ e^{iqz}.
\label{13}
\end{equation}
Assuming $\Delta (z)$ is analytic at $z=0$, we may write
\begin{equation}
  \label{expmom}
  \Delta (z) = \sum _ {n=0} ^{\infty} {(-1)^n \over (2n)!} ~ z^{2n}~
  Q_{2n},
\end{equation}
where
\begin{equation}
  Q_{2n}\equiv \int _q ~ q^{2n} ~ \Delta _q
\label{moments}
\end{equation}
will be called moments of the noise correlator $\Delta$.

Equation (\ref{10}) can now be solved by iteration.
As an example, we insert the second term $\tilde G_0 v^{-1} \tilde
\eta$ on the $rhs$ of Eq. (\ref{10}) for the $\tilde h$ of the fourth
term which gives a ``one-loop'' contribution to the solution $ \tilde
h({\bf k}, \omega)$. Neglecting higher order terms and averaging Eq.
(\ref{10}) over the disorder, we obtain (see Fig. 4)
\begin{equation}
\label{intexa}
v\simeq {1\over\lambda}F+ {1 \over \lambda} \int _q \int _{\bf p}
{}~{iq~\Delta _q \over \gamma {\bf p}^2 - i\lambda vq }.
\end{equation}
For $D>D_c =4$, the integral in the correction term converges for all
$v$ (see Appendix A.1).

When $D<4$, the integral in (\ref{intexa}) diverges in the limit
$v\rightarrow 0$ due to contributions from small momenta, $p \to 0$
(see Appendix A.1).  This statement can be confirmed by continuing the
iteration to higher orders (see Appendix A.2).  Quite generally, the
expansion parameter can be written in the form $(\xi _0/L_c)^{4-D}$,
where
\begin{equation}
  \xi_0 = (\gamma a_{\perp} / \lambda v) ^{1/2}
\end{equation}
plays the role of a bare correlation length.  This parameter is small
at high velocities.

\subsection{Linear response
and renormalization of $\gamma$ and $\lambda$}

In the moving phase, the interface is free on large length and time scales.
The dynamics of the system for small $k$ can thus be described
by linear response theory with effective parameters.
As usual, one can
define an effective linear response function
$\tilde G({\bf k},\omega)$ \cite{HohHal},
\begin{equation}
\label{11}
\tilde G({\bf k},\omega)= \left. \frac{\partial \tilde h({\bf k},\omega)}
{\partial \tilde \varphi ({\bf k},\omega)}
\right |_{\tilde\varphi=0}.
\end{equation}
Due to the nonlinear terms in Eq. (\ref{10}),
$ \tilde G({\bf k},\omega) $ differs from
$\tilde G_0 ({\bf k},\omega)$.
The correction can be expressed in terms of a ``self-energy''
$\tilde \Sigma ({\bf k}, \omega)$,
\begin{equation}
\label{15}
\tilde G ({\bf k}, \omega) =
\tilde G_0 ({\bf k}, \omega) + \tilde G_0 ({\bf k}, \omega)
\tilde \Sigma ({\bf k}, \omega) \tilde G ({\bf k}, \omega).
\end{equation}
As we shall show, $\tilde \Sigma ({\bf k}, \omega)$
has the same functional form as
$\tilde G_0^{-1} ({\bf k}, \omega)$ at small ${\bf k}$ and $\omega$.
Hence the correction term in (\ref{15}) can be absorbed into
the free one with the bare parameters $\gamma $ and $\lambda$
replaced by effective parameters
$ \lambda _{eff} = \lambda + \delta \lambda $ and
$ \gamma _{eff} = \gamma + \delta \gamma $.

To calculate $\delta \gamma$ and
$\delta \lambda $ one has to consider diagrams with one incoming
and one outgoing line (see Eq. (\ref{15})).
The lowest order contribution to the
self energy $\tilde \Sigma ({\bf k}, \omega)  $
is given by
the second and third diagram on the $rhs$
of the diagrammatic perturbation expansion for $ \tilde G({\bf k},\omega) $
shown in Fig. 5.
The corresponding analytic expression is
\begin{equation}
\tilde \Sigma ({\bf k}, \omega) \simeq
\int _q \int _{\bf p } \Delta _q
\label{16}
\left \{ { (iq)^2 \over
\gamma {\bf p}^2 + i \lambda qv }
- { (iq)^2 \over
\gamma ( {\bf p+k})^2 + i \lambda ( \omega + qv )}
\right \}.
\end{equation}
For small ${\bf k}$ and $\omega$
Eq. (\ref{16}) can be written as
\begin{equation}
\tilde \Sigma ({\bf k}, \omega) \simeq
\left [ \gamma {\bf k}^2\left ({4 \over D} -1 \right )
- i \lambda \omega \right ]
\int _q \int _{\bf p }   { q^2 \Delta _q \over
(\gamma {\bf p}^2 + i \lambda vq)^2 }.
\end{equation}
As in Eq. (\ref{intexa}),
the integration over $p$ converges at large $p$
as long as $ 4-D = \epsilon > 0 $, while
$\sqrt {\lambda v q / \gamma} $ serves as a lower cutoff for $p$.
The $q$ integration is truncated for large $q$ by the correlator
$\Delta _q$ which is assumed to decay as $e^{-q^2a^2_{\bot}}$.
The integrals can be evaluated to give
\begin{equation}
\tilde \Sigma ({\bf k}, \omega) \simeq
Q_2 ~{K_D \over \gamma ^2} ~{ 1 \over \epsilon} \left ( {\gamma a_{\bot}
\over \lambda v}
\right)^{\epsilon /2}
\left [ \gamma {\epsilon \over D } {\bf k} ^2 - i \lambda \omega \right]
\end{equation}
(see Appendix A.1, where similar integrals are carried out).
When comparing this with Eq. (\ref{10a}) and using (\ref{15}),
the first order corrections
$ \delta \gamma ^{(1)} $ and $ \delta \lambda ^{(1)} $
to $ \gamma $ and $ \lambda $ are \cite{gammaren}
\begin{eqnarray}
\label{17a}
\delta \gamma ^{(1)} &=& - \gamma c Q_2 {1\over D} \xi_0^\epsilon \equiv
-\gamma {K_D \over D} \left ( {\xi_0 \over L_c} \right ) ^\epsilon , \\
\delta \lambda ^{(1)} &=&  \lambda c Q_2 {1 \over \epsilon} \xi_0 ^\epsilon
=\lambda {K_D \over \epsilon} \left ( {\xi_0 \over L_c} \right ) ^\epsilon ,
\label{17b}
\end{eqnarray}
where $c \equiv K_D /\gamma ^2$ and $L_c^{\epsilon}=\gamma^2/Q_2$.
With these notations the small parameter of the perturbation theory
can be written as $c Q_2 \xi_0^\epsilon$ or as
$(\xi_0/L_c)^\epsilon$. We see that $Q_2 = -\Delta ''(0)$ plays the
role of a coupling constant.

The result for $ \delta \lambda ^{(1)} $ in (\ref{17b})
can also be obtained by
considering the lowest order correction
Eq. (\ref{intexa}) to the bare velocity $ v_0 = F/\lambda $,
as done by Feigel'man \cite{Feigelman}.
In Appendix A.2 the calculation of the velocity
is extended to second order,
with the result,
\begin{equation}
\label{velsec}
\lambda _ {eff} = \lambda \left[1 +
{\delta \lambda ^{(1)} \over \lambda}+
\left({\delta \lambda ^{(1)} \over \lambda}\right)^2+...\right].
\end{equation}

The series obtained by perturbation theory
are thus divergent for $ v \to 0 $ and cannot be used directly close to
the depinning transition.

\subsection{Interface width}

The width of the interface $ \langle h^2 ({\bf x} , t) \rangle$
can be calculated from Eq. (\ref{10}) in a very
similar way as the velocity.
The lowest order diagram for the width is shown in
Fig. 6. The corresponding analytical expression is given by
\begin{equation}
\langle h^2 ({\bf x} , t) \rangle =
\int _q
\int _{\bf p} ~\Delta _q~
 { 1 \over
\gamma ^2 {\bf p}^4 +  (\lambda vq)^2 }
\end{equation}
which can be evaluated easily, giving
\begin{eqnarray}
\label{22}
\langle h^2 ({\bf x} , t) \rangle &\simeq &
a_\perp^2 ~ {2 K_D \over D-2} ~ {1 \over \epsilon}
{\Delta (0) \over a_\perp^2 \gamma ^2}
\left ({a_\perp \gamma \over \lambda v} \right) ^{\epsilon/2}  ~
+ O(\epsilon) \\ \nonumber
&\simeq &
a_\perp^2 ~ {2 K_D \over D-2} ~ {1 \over \epsilon}
{ \left ( \xi_0 \over L_c \right ) ^\epsilon } ~+ O(\epsilon) .
\end{eqnarray}
The interface width diverges with $v \to 0$. Since $\xi_0$ is
a correlation length, one obtains from Eq. (\ref{22})
the perturbative roughness exponent $\zeta _{p} = \epsilon /2$.

It can be shown that the second order contribution to
$ \langle h^2 ({\bf x} , t) \rangle $ is proportional to the square of
the first order Eq. (\ref{22}) but with one factor $ 1/\epsilon $ lacking:
\begin{equation}
\label{27}
 a_\perp^2 ~ \left ( {2 K_D \over D-2} \right )^2~ {1 \over \epsilon}~
 \left ( {\xi_0 \over L_c }\right ) ^{2 \epsilon}  .
\end{equation}
Here we have implicitly assumed that $ \Delta '(0) = 0$ which will
be discussed below (Sec. V.).

\subsection{Renormalization of the moments of $\Delta(z)$}

In the preceding section we have considered the renormalization of $\gamma $
and $\lambda $. Eqs. (\ref{17a}-\ref{17b}) show that the parameter
controlling the renormalization of $\lambda $ is proportional to the second
moment of the disorder correlator $Q_2=\int_qq^2\Delta _q$. In order to
carry out the renormalization group analysis of the problem it is necessary
to know the renormalization prescription of $Q_2$. The usual way to get this
prescription is to find a vertex function,
which is proportional to $Q_2$ when calculating the
lowest-order contribution in the disorder strength.
Then, the
higher-order contributions to this vertex function yield the desired
renormalization prescription of $Q_2$. It turns out that a convenient quantity
to look at for this purpose is

\begin{eqnarray}
G_{11}=&&\int_{\Omega _1}\int_{\Omega _1^{^{\prime }}}
\tilde G_0^{-1}({\bf k},\omega )
\tilde{G}_0^{-1}({\bf k}^{\prime },\omega^{\prime })
\nonumber\\
&&\times\Biggl \langle \left. {\frac{\delta \tilde{h}({\bf k},\omega )}
{\delta \tilde{h}_0({\bf p}_1,\Omega_1)}}
\right|_{\scriptstyle \tilde{h}_0=0,\omega =0\atop
\scriptstyle{\bf k}={\bf p}_1={\bf 0}}{}
\left. {\frac{\delta \tilde{h}({\bf k}^{\prime},\omega ^{\prime })}
{\delta \tilde{h}_0({\bf p}_1^{\prime},\Omega_1^{\prime})}}
\right|_{\scriptstyle \tilde{h}_0=0, \omega^{\prime}=0\atop
\scriptstyle{\bf k}^{\prime }={\bf p}_1^{\prime }={\bf 0}}\Biggr\rangle_c,
\label{q_2}
\end{eqnarray}
where $\tilde{h}_0({\bf p},\Omega )=\tilde{G}_0({\bf p},\Omega )\tilde{%
\varphi}({\bf p},\Omega )$.
The subscript $c$ indicates that only connected diagrams
are to be considered. The perturbation expansion of (\ref{q_2}) in powers
of the disorder correlator is obtained by iterating $\tilde{h}({\bf k}%
,\omega )$ in Eq. (\ref{q_2}) according to Eq. (\ref{10}) and carrying out the
average over the disorder. This perturbation expansion can be represented by
means of diagrams. The lowest order contribution to (\ref{q_2}) in powers of
$\Delta (z)$ is found by replacing all $\tilde{h}$ on the $rhs$ of Eq. (\ref
{10}) by $\tilde{h}_0$. The first-order contribution to (\ref{q_2}) is given
by diagram a) in Fig. 7 with $n=1$. According to Eq. (\ref{q_2}) and
Eq. (\ref{10}) the analytical expression
associated with the diagram 7a)
is exactly the
second moment $Q_2$. The renormalization prescription of $Q_2$ can be
obtained from the one-loop diagrams in Fig. 7b)-d) by setting $n=1$.
In each of these
diagrams, there is only one integration over the internal momentum
${\bf p}$.
This
integration has the form $\int_{\bf{p}}G_0^2({\bf p})$ and yields for
all diagrams under consideration the same factor $c\xi _0^\epsilon /\epsilon
$. In addition to this integration, there is also a factor which involves
an integration over the
momenta $q_1$ and $q_2$ conjugated to the argument of the disorder
correlator $\Delta (z)$.
This factor results in
\begin{equation}
\int_{q_1}\int_{q_2}(-q_1q_2^3+3q_1^2q_2^2-3q_1^3q_2)\Delta (q_1)\Delta
(q_2)=3Q_2^2,  \label{q_2,1}
\end{equation}
where we have taken into account that only even moments of the disorder
correlator differ from zero. The total contribution of the diagrams in
Fig. 7 for $n=1$ leads to the following renormalization prescription of
$Q_2$,
\begin{equation}
Q_{2,eff}=Q_2(1+{3Q_2c{\frac 1\epsilon }\xi _0^\epsilon +...).}
\label{q_2,2}
\end{equation}

A detailed analysis of the present problem shows that, besides the
renormalization of the second moment, higher moments $Q_{2n}$ ($n=2,...$)
also renormalize. The renormalization of $Q_{2n}$ can be
derived by considering the following generalization of the expression (\ref
{q_2}),
\begin{eqnarray}
\label{verana}
G_{nn}=&&\int_{\Omega _1}\int_{\Omega _1^{^{\prime }}}
\tilde G_0^{-1}({\bf k},\omega )
\tilde{G}_0^{-1}({\bf k}^{\prime },\omega^{\prime })
\nonumber  \\
&&\times\Biggl \langle \left. {\frac{\delta^n \tilde{h}({\bf k},\omega )}
{\delta \tilde{h}_0({\bf p}_1,\Omega_1)
\delta \tilde{h}_0({\bf p}_2,\Omega_2)\cdots
\delta \tilde{h}_0({\bf p}_n,\Omega_n)}}
\right|_{\scriptstyle \tilde{h}_0=0,\omega =\Omega_2=\ldots=\Omega_n=0\atop
\scriptstyle{\bf k}={\bf p}_1=\ldots={\bf p}_n={\bf 0}}{}
\\
\nonumber
&&\qquad\left. {\frac{\delta^n\tilde{h}({\bf k}^{\prime},\omega ^{\prime })}
{\delta\tilde{h}_0({\bf p}_1^{\prime},\Omega_1^{\prime})
\delta\tilde{h}_0({\bf p}_2^{\prime},\Omega_2^{\prime})\cdots
\delta\tilde{h}_0({\bf p}_n^{\prime},\Omega_n^{\prime})
}}
\right|_{\scriptstyle\tilde{h}_0=0,\omega'=\Omega_2'=\ldots=\Omega'_n=0
\atop
\scriptstyle{\bf k}'={\bf p}_1'=\ldots={\bf p}_n'={\bf 0}}\Biggr\rangle_c.
\end{eqnarray}
The perturbation expansion of (\ref{verana})
is obtained in the same way as the iteration of Eq. (\ref{q_2}).
The lowest-order contribution to
(\ref{verana}) in powers of $\Delta (z)$ is
found by replacing all $\tilde{h}$ on
the $rhs$ of Eq. (\ref{10}) by $\tilde{h}_0$. Only the diagram with $n$
outgoing lines and one dotted line in Fig. 3 gives a contribution to the
first-order correction of $G_{nn}$. As it can be easily shown,
the latter is
equal to the bare moment $Q_{2n}$ (see Eq. (\ref{expmom}))
and is represented by
the diagram a) in Fig. 7. The second-order
contributions to $G_{nn}$ in powers of $\Delta $ are given by diagrams b)-d)
in Fig. 7. The renormalization of the moments $Q_{2n}$ and consequently
the renormalization of the disorder correlator $\Delta (z)$, can be obtained
by computing $G_{nn}$ up to the second order in $\Delta $. The details of
such one-loop computation of $G_{nn}$ are presented in Appendix B. As a
result, the renormalization prescription of the moments $Q_{2n}$ is obtained
as {\
\begin{equation}
Q_{2n,eff}-Q_{2n}\simeq \delta Q_{2n}^{(1)}\simeq
\sum_{k=1}^nC_{2k}^{2n+1}Q_{2k}Q_{2n-2k+2}~{\frac c\epsilon }\xi _0^\epsilon
,  \label{28}
\end{equation}
where $C_i^j={\bigl (}{_i^j}\bigr ) $ are the binomial coefficients (see
Appendix B). In particular, the correction to the second moment $Q_2$ is
given by Eq. (\ref{q_2,2}).  Eq. (\ref{28}) has been derived for }$n=1,2,...$.
However, it can be extended to $n=0$. Taking into account that $C_2^1=0$ we
get from (\ref{28}) that the zeroth moment $Q_0$ does not renormalize. This
conclusion is in agreement with the result on the
interface width in the preceding section. Equation (\ref{28}) will be used in
Sec. IV to derive the flow equation for the
correlator $\Delta (z)$.

\section {Renormalization group}

Approaching the critical point with $v \to 0$,
the perturbative corrections to the parameter of the model
diverge as $v^{-\epsilon /2}$, which is caused by the integration
over small momenta. A possible way to sum up such singularities
is to perform a renormalization group (RG) procedure, where one integrates
out fluctuations in an infinitesimal momentum shell only,
leading to differential flow equations. From the flow of the relevant
parameters the critical behavior can be deduced. The calculation is
simplified close to the upper critical dimension $D_c = 4$.
We will therefore
determine the critical exponents of the depinning transition
by a RG procedure in $D=4-\epsilon$ dimensions.

\subsection{Three-parameter RG}

The smallest closed set of flow equations
to first order in $\epsilon$
is given by \cite{NSTL}
\begin{eqnarray}
\label{30a}
d\ln\gamma/d\ln L&=&0, \\
\label{30b}
d\ln\lambda/d\ln L&=&cQ_2L^\epsilon, \\
\label{30c}
dQ_2/d\ln L&=&3cQ_2^2L^\epsilon
\end{eqnarray}
($c \equiv K_D/\gamma ^2$).
To derive these flow equations we consider
$ \gamma _{eff}$, $\lambda _{eff}$ and
$Q_{2,eff}$ as depending
on the upper cut-off length, $L=\xi_0$.
Eqs. (\ref{30a})-(\ref{30c})
can be formally obtained by differentiating
Eqs. (\ref{17a}), (\ref{17b}), and (\ref{q_2,2})
and then setting $ \gamma _{eff} = \gamma $,
$ \lambda_{eff} = \lambda $, and $ Q_{2,eff} = Q_2 $.
This is equivalent to the RG transformation, where the degrees of freedom
in a momentum shell are successively integrated out.
There, the corrections to
the parameters are calculated in the same way as in perturbation theory -
the only difference is that one integrates only over the momentum shell
instead over all momenta.

Note that the perturbative correction (\ref{17a}) for $\gamma$ is {\it finite}
in the limit $\epsilon \to 0$ and does not contribute to first order
to a renormalization of $\gamma$ \cite{gammaren}.
The flow equation (\ref{30c}) can also be obtained from the second order
correction (\ref{velsec}) to the velocity \cite{Stepanowann}.

Since there is no $1/\epsilon ^2$ term in the second order correction
(\ref{27}) of the interface width, provided $\Delta '(0)=0$, we conclude that
$\Delta (0) $ does not renormalize to first order in $\epsilon$.

When integrating Eq. (\ref{30c}) we see that
$Q_2$ becomes infinite at $L \simeq L_c$
(if we assume that
$\Delta (z)$ is analytic and monotonously decaying).
It then follows from Eq. (\ref{30b}) that also $\lambda $ diverges
at $L\simeq L_c$. This means that beyond $L_c$, there is no dynamic response
to any finite force, which is an unphysical result !

\subsection{Functional RG}

Eq. (\ref{28}) shows that
besides the renormalization of $Q_2$ there is also a renormalization of
all the other moments $Q_{2n} (n=2,...)$.
The renormalization of the moments causes
a renormalization of the disorder correlator
$\Delta (z)$\cite{Stepanowann}.
Since all $Q_{2n}$ have the same
upper critical dimension we can put together
the prescriptions (\ref{28}) into a
functional flow equation \cite {Stepanowann} for $ \Delta (z) $
(see Appendix C):
\begin{equation}
\label{30d}
{d~\Delta (z) \over d~\ln L } = - c L^{\epsilon}
{d^2 \over d z^2 } \left [ {1 \over 2} \Delta^2 (z)
- \Delta (z) \Delta (0) \right ].
\end{equation}

To discuss the fixed point behavior it is convenient to consider
the renormalized {\it and} rescaled correlator $\hat \Delta (z)$,
which has the original short-distance cutoff of the bare correlator.
To do so we
perform a scale transformation ${\bf x}\rightarrow L{\bf x}$,
$t\rightarrow L^z t$, and $h\rightarrow L^\zeta h$ with Eq. (\ref{7}).
At the critical point we expect scale invariance.
Since $\gamma$ is not renormalized, the renormalized and rescaled
equation of motion,
\begin{equation}
\label{36}
\lambda L^{2-z} {\partial h \over \partial t} = \gamma
\nabla ^2 h + L^{2-\zeta}(F - \lambda v)
+L^{2-\zeta} \eta \bigl(L{\bf x} , vL^z t +L^\zeta h\bigr),
\end{equation}
should be independent of $L$. This is the case when
$\lambda(L)\sim L^{z-2}$ and
\begin{equation}
\label{32}
\Delta(z)={\rm const.} \times L^{2\zeta-\epsilon}
\hat \Delta(zL^{-\zeta}).
\end{equation}

To investigate the fixed point solution
$\Delta^\ast(y)=\lim_{L\rightarrow\infty} \hat \Delta (y)$
we insert (\ref{32}) into (\ref{30d}), take the limit
$L\rightarrow\infty$, and determine the constant in Eq. (\ref{32}) such
that $\Delta^\ast(0)=1$,
\begin{equation}
\label{33}
(\epsilon-2\zeta)\Delta^\ast(y)+\zeta y{\Delta^\ast}'(y)
-[{\Delta^\ast}'(y)]^2-{\Delta^\ast}''(y)[\Delta^\ast(y)-1]=0.
\end{equation}
The same equation with
$\Delta ^\ast (y) = R^{\prime \prime \ast}(y)$
was already derived for the potential-potential correlator $R(y)$
in the problem of an equilibrium interface subject to a random
potential \cite{Fisherequ}. From that treatment
it is known \cite{Fisherequ} that the
fixed-point function $\Delta ^\ast (y)$ has a singularity
at the origin. The ansatz
\begin{equation}
\label{34}
\Delta^\ast(y)=1+a_1|y|+{1\over 2}a_2y^2+\ldots
\end{equation}
satisfies Eq. (\ref{33}) and by
a comparison of the coefficients of $y^0$ and $y^1$
one gets $a_1^2=\epsilon-2\zeta$ and $a_2=(\epsilon-\zeta)/3$.

The development of this cusp singularity can be
understood as follows (see Fig. 8).
Starting the renormalization procedure at $L=L_0$
with a smooth correlator ($\Delta ''(0) <0$),
the curvature of $\Delta (z) $ at the origin becomes greater
under
the flow (\ref{30c}). Hence, the position $z_0$ of the point where the
second derivative $\Delta ''(z)$ changes its sign is shifted
closer to the origin. A diverging curvature means $z_0 = 0$, i.e.
$\Delta ''(z \to 0^+) > 0$.

Repeating the perturbative calculations with a correlator obeying
Eq. (\ref{34}) we find that Eqs. (\ref{17a}), (\ref{17b})
and (\ref{30a}), (\ref{30b}) remain valid to
first order in $\epsilon$.
With the interpretation
$Q_2=\Delta''(0^+)>0$ we overcome the above unphysical result
that the inverse mobility diverges at $L=L_c$.

Since our earlier assumption $\Delta'(0)=0$ failed,
which yields (\ref{27}), the conclusion of the nonrenormalization of
$ \Delta (0) $ is no longer valid. Instead we obtain with Eq. (\ref{34})
$ \partial \Delta (0) / \partial \ln L=- c L^\epsilon [{\Delta}'(0)]^2$
which is in agreement with Eq. (\ref{33}) for $y=0$ when using Eq. (\ref{32}).

The singular term $|z|$ yields a reduction of
the driving force, $F\rightarrow F-F_c$.
To see this we split the correlator $\Delta$ in a regular and a singular
part. In Fourier-space, the singular part behaves as
$\Delta _q ^{\rm sing} \sim - q^{-2} \Delta '(z \to 0^+)$. Using this
in Eq. (\ref{intexa}) for the velocity, we obtain
\begin{equation}
F_c \simeq -   {K_D \over D-2}{1 \over \gamma}
\Delta'(L_0,z \to 0^+)\Lambda_0^{D-2},
\end{equation}
which depends on the upper cutoff $\Lambda_0\simeq \pi/L_0$ of
the momentum space integration.
Using $\Delta '(0^+) = \Delta (0) /a_\perp $ (see Eq. (\ref{34})) and
identifying $L_0$ with $L_c$ yields the threshold  $F_c$
of Eq. (\ref{F_c}), in agreement with the estimate
of the scaling arguments \cite{Larkin,Feigelman,BruAep}.

\subsection{Critical exponents}
The roughness exponent $\zeta$ can be obtained immediately from
Eq. (\ref{33}), provided $\Delta ^\ast (y) $ goes sufficiently
fast to zero with $y \to 0$. Integration over $y$ yields
$(\epsilon - 3 \zeta ) \int _{-\infty} ^\infty dy~ \Delta ^\ast (y)=0$.
For random-field disorder
$\int _{-\infty} ^\infty dy~ \Delta ^\ast (y) > 0$, from which one gets
\begin{equation}
\zeta = \epsilon / 3.
\end{equation}

The other exponents can be read off from Eq. (\ref{36}).
To calculate the dynamical exponent $z$ we have to know the
scale dependence of $\lambda$, which can be obtained by
integrating (\ref{30b}) using (\ref{32}) and (\ref{34}),
\begin{equation}
\label{35}
\lambda(L)=\lambda_0(L/L_0)^{-(\epsilon-\zeta)/3},
\end{equation}
where $\lambda_0=\lambda(L_0)$.
Using Eqs. (\ref{mcoreta}), (\ref{32}),
(\ref{35}) and imposing scale invariance of Eq.  (\ref{36})
at $F \to F - F_c = v=0$, $z$ is given by
\begin{equation}
\label{37}
z=2-(\epsilon-\zeta)/3.
\end{equation}

As discussed earlier,
above the threshold the quenched
character of the noise correlator changes to a thermal one
on length scales $L\ge\xi \sim v^{-1/(z-\zeta)}$.
Hence we have to stop
the renormalization at $\xi$. From $F-F_c=\lambda(L=\xi) v$ follows
\begin{eqnarray}
\label{38a}
v\sim (F-F_c)^\theta,\qquad &{\rm with}& \quad \theta=1-{1\over 3}
{\epsilon-\zeta\over 2-\zeta} \\
\label{38b}
\xi\sim (F-F_c)^{-\nu},\qquad &{\rm with}& \quad \nu=1/(2-\zeta)
\end{eqnarray}
where we have used Eqs. (\ref{35}) and (\ref{37}).

Inserting the result for random-field disorder
$\zeta=\epsilon/3$ in Eqs. (\ref{37}) and (\ref{38a}),
(\ref{38b})
we obtain the final results for the critical exponents to
first order in $\epsilon$,
$$\hskip 0.8truecm \zeta\simeq{1\over 3}\epsilon,
\hskip 1.55truecm z\simeq 2-{2\over 9}\epsilon,$$
\begin{equation}
\label{41}
\theta \simeq 1 - {1\over 3 } {2 \epsilon \over 6 - \epsilon}
\simeq 1-{1\over 9}\epsilon,\qquad
\nu\simeq {3 \over 6 - \epsilon} \simeq{1\over 2}+{1\over 12}\epsilon.
\end{equation}

The roughness exponent $\zeta=\epsilon/3+O(\epsilon^2)$
differs from the value $\zeta _p = \epsilon/2$ from perturbation theory
(see Sec. III. C) \cite {EfeLar,Parisi}, but coincides with
that of an equilibrium interface in a medium with random-field disorder
\cite{NatRuj}.
Recently,  Narayan and Fisher \cite {NarFisint} showed that
higher order terms in Eq. (\ref{30d}) vanish when integrated over all $z$.
Thus, the results for the exponents $\zeta $ and $\nu$ could
be exact, although non-perturbative corrections cannot be ruled out
\cite {NarFisint}.
In fact, as we will see in the next section, our simulations
in $D \le 3$ yield values for $\zeta $ and $\nu$
which differ slightly from the results (\ref{41}).

The velocity exponent $\theta $ is smaller than one for $D <  4$ and
approaches the mean field value $\theta _{\rm MF} = 1$ for $D=4$
(see
Sec. II). The superdiffusive dynamical exponent $z<2$ is consistent
with the expectation that the motion of the interface close to threshold
is governed by the behavior of avalanches \cite{BTW}.
The Harris condition (\ref{harcon})
is fulfilled as an equality to the first
order in $\epsilon$.
Narayan and Fisher \cite{NarFisint} showed that the scaling relation
(\ref{38b}) between $\zeta $ and $\nu$ is exact due to a
symmetry of Eq. (\ref{8}).
Thus, the Harris condition (\ref{harcon})
can be written as $\zeta \ge (4-D) /3$.

Narayan and Fisher \cite{NarFisint} investigated also the RB case
(see Sec. II) and argued that the results (\ref{41})
are also valid for RB disorder.
The critical exponents for CDW's can be obtained by inserting the
corresponding value $\zeta _{CDW} = 0$ into the scaling relations
(\ref{37}), (\ref{38a}) and (\ref{38b}) which are independent of the
form of the correlator $\Delta (z)$ (see Sec II.) \cite{NarFiscdw}.

\subsection{Phase diagram}

The crossover behavior of the equation of motion (\ref{8}) can be
summarized in a $F-L$ phase diagram which is shown in Fig. 9. On small
scales $L<L_c$ the height fluctuations are bounded by the microscopic
correlation length $a_{\bot}$ of the random forces.  Here, the correlations
(\ref{mcoreta}) can be described by a smooth function $\Delta (z)$.  On
larger length scales $L_c < L < \xi$ the system is in a critical state.
The behavior of the interface is characterized by the critical exponents
of the depinning transition.  When crossing the Larkin length $L_c$, the
second derivative $\Delta ''(0)$ diverges under the
RG flow and the fixed-point
function $\Delta ^\ast (z) $ of the correlator develops a cusp singularity at
the origin.  When going to length scales larger than $\xi$
(for $F>F_c$), the
behavior crosses over to the EW regime \cite{EW}.  The full equation of
motion (\ref{meqmsqg}) has (for $T=0$)
an additional regime, namely the one of the
KPZ equation \cite{KPZ}.  For $D>1$ this regime is situated above the
region of the EW equation.

\section{Numerical results}

In this section we compare the analytical results for the critical
exponents with extensive numerical simulations \cite{gimo,PhD}.
Since our analytical values for the exponents (\ref{41}) have been
obtained by an expansion to first order in $\epsilon$, it is an
open question whether or not the results (\ref{41}) are
applicable to $\epsilon \ge 1$.
However,
further work by Narayan and Fisher \cite {NarFisint} suggests that the
first-order results for the exponents $\zeta $ and $\nu$ are exact.
We will see below that this is not
consistent with our numerical simulations. The deviations between the
analytical results and the simulations are much larger than the error
bars on the numerical values but decrease with increasing $D=4-\epsilon$,
thereby supporting our analytical calculation.

\subsection{Simulation of the continuum equation in $D=1$}

We first simulate Eq. (\ref{8}) in $D=1$ with a discretization of the
transverse coordinate only, $x \to i,~z(x,t) \to z_i(t)$.
The random forces $\eta _i (z_i)$ are chosen as follows:
Each integer position $z_i$ on a square lattice
is assigned a random number $\eta$ between zero and one.
For non-integer $z_i$ the forces $\eta _i (z_i)$ are obtained
by linear interpolation \cite{Parisi}.
Finally, the $z$-coordinates of $\eta _i (z_i)$
in each column $i$
are shifted by a random amount $0 \le s_i < 1$, i.e.,
$\eta _i(z_i) \to \eta _i (z_i+s_i)$ \cite{PhD}.

At $t=0$ the interface is flat, $z_i(t=0) \equiv 0$.
The interface configuration at $t+\Delta t$ is calculated
simultaneously for all $i$ using the method of finite differences,
\begin{equation}
\label{4eulalg}
z_i (t +\Delta t) = z_i (t) + \Delta t \Big [
F + z_{i+1} (t) + z_{i-1}(t)  - 2z_i (t) + m \eta _i (z_i) \Big ].
\end{equation}
Here $m$ is a real parameter and periodic boundary conditions are used.

We are interested in the critical exponents, so the threshold
$F_c$ has to be determined from a measurement of the velocity $v(F)$
close to $F_c$ where Eq. (\ref{defthe}) is valid.
However, it turns out that the critical region is very narrow and
that an accurate determination of $v(F)$ inside that region is
difficult, which seems to be the reason for contradictory results
for $\theta $ in the literature \cite{Parisi,Csahok,DonMar}.
At the beginning of the interface motion the velocity decreases and only
saturates after a time $t_v$ when
the roughness $w \sim \xi ^\zeta $ no longer increases.
The time $t_v$ behaves as $t_v \sim (F-F_c)^{-z\nu}$
and therefore the simulations become increasingly time-consuming
when approaching the threshold. In addition, the average velocity $v$
is systematically underestimated if the temporal fluctuations of the
velocity become comparable with $v$ itself. Thus, for $F$ close to $F_c$
very large systems have to be simulated.

For $m=3$ we estimated the threshold to be in the interval
$1.41 \le F_c \le 1.42 $ and
for the velocity exponent we obtained
an upper bound, $\theta (D=1) \le 0.45$.

To determine the roughening exponent $\beta \equiv \zeta /z$,
we measured the interface width
$w^2 (L,t) = \Bigl \langle \overline {[z_i(t) - \overline
{ z_i(t)}]^2 } \Bigr \rangle$ and the average position
$H(t) \equiv \langle \overline {z} (t) \rangle $
in the transient regime, $t \ll L^z,\xi ^z$
for various forces in the interval $1.41 \le F \le 1.42 $.
($L$ is the linear system size,
the overbar denotes the spatial average over $i$,
and the angular brackets stands for the average over
configurations of the disorder.)
The best scaling of the width,
$w^2 \sim t^{2 \beta}$, is achieved at $F=F_c \simeq 1.4135$,
with $\beta = 0.88 \pm 0.03$, which is shown in Fig. 10.
The average position $H(t)$ increases with an effective exponent which is
only slightly lower, $\beta _H \simeq 0.85$.

A possibility to determine both, $\beta $ and the roughness exponent
$ \zeta $ is to consider
the height-height correlation function
$ C(r,t)=\langle \overline {[z_{i+r}(t) -  z_i (t)]^2}   \rangle$.
In the transient regime it scales as
\begin{equation}
\label{scaC(r}
C (r,t) \sim r^{2\zeta} \Psi ( r t^{-1/z}),
\end{equation}
where $\Psi (y)$ is a scaling function.
Thus, the exponents $\zeta $ and $\beta$ can be obtained
by a scaling plot of $C(r,t)$ which is shown in Fig. 11.
The best data collapse at $F=F_c$ is achieved with
$\zeta (D=1) = z\beta = 1.25 \pm 0.05$ and $\beta (D=1) = 0.88 \pm 0.03$.

\subsection{Simulation of the automaton model}

Since the simulations of the continuum equation (\ref{4eulalg}) are
computationally expensive,
we also study a lattice model \cite{gimo} of probabilistic cellular
automata, which allows to determine effectively
the critical exponents in higher dimensions as well.

The automaton model is defined on a $D+1$ dimensional hypercubic lattice
where each cell $[i,z]$ (with $1 \le i \le L^D$) is assigned a random
force $ \eta _{i,z}$ which takes the value 1 with probability $p$
and $ \eta _{i,z} = -1 $ with probability $1-p$.
During the motion for a given time $t$ the local force
\begin{equation}
\label{43}
f_i (t) = \sum_{ \langle j \rangle} \bigl [ z_{j} (t) -  z_{i} (t) \bigr ]
 + m \eta_{i,z_i}
\end{equation}
is determined for all $i$, where the sum is over nearest neighbors
$ \langle j \rangle $ only and
$m$ is an integer parameter.
The interface configuration is then updated  simultaneously for all $i$:
\begin{eqnarray}
z_i (t+1) &=& z_i (t) +1 ~~~~~~~ {\rm if} ~~v_i > 0 \nonumber  \\
z_i (t+1) &=& z_i (t) ~~~~~~~~~~~~{\rm otherwise.}
\label{44}
\end{eqnarray}
The parameter $m$ measures the strength of the random force compared to the
elastic force and
the difference $p-(1-p)=2p-1$
determines the driving force.
The growth rule specified by Eqs. (\ref{43}) and (\ref{44})
can be derived from the continuum equation (\ref{8}) by discretizing
the transverse directions ${\bf x}$, as well as time and the position
of the interface $z({\bf x},t)$. In addition,
a simple two-state random force $\eta_{i,z} =\pm 1$
is used which contributes to
the computational advantage that one has to handle with integer
variables only.

\smallskip


The exponents $\zeta $ and $\beta $ are obtained in the same
way as in Sec. V.A \cite{gimo}.
In addition, the static roughness exponent $\zeta $ is determined
by measuring the interface width
of pinned interfaces for various system sizes $L$.
The scaling $w^2 \sim L^{2 \zeta }$ at threshold $p=p_c$
is shown in Fig. 12 ($D=$1,2 and 3).
In $D=1$ we obtain the same results for $\zeta $ and $\beta $ as for
Eq. (\ref{4eulalg}) but with smaller error bars.
This supports the expectation that the continuum equation (\ref{8}) and the
automaton model belong to the same universality class.

The results for $\zeta $ and $\beta $ are shown in Table 1, together
with the other critical exponents $z \equiv \zeta /\beta $, $\theta$,
and $\nu$.
The correlation length exponent $\nu$ is calculated from the scaling
relation (\ref{38b}) which was shown to be exact \cite{NarFisint}.
(Details can be found in Refs. \cite{gimo} and \cite{PhD}).
The velocity exponent $\theta $
is determined from
the data of Fig. 13 (see Table 1).

\subsection{Self-organized automaton model (SOAM)}

The difficulty of the simulations
of Eq. (\ref{4eulalg}) and of the automaton model that
the critical region is narrow
can be circumvented by ``self-organizing''
the automaton model (\ref{43}), (\ref{44}). Then, the system is always
at threshold without tuning a parameter \cite{BTW}. This can be achieved
by using an idea of Havlin et al. \cite{Havlin} and Sneppen \cite{Sneppen}:
Instead of updating the whole interface configuration simultaneously
as in Eq. (\ref{44}) one increments only that column $j$ where
the local force (\ref{43}) takes the maximum
$f_{\rm max} \equiv \max _i [f_i]$ among all $f_i$ \cite{PhD},
\begin{eqnarray}
z_i (t+1) &=& z_i (t) +1 ~~~~~~~ {\rm if} ~f_i= f_{\rm max} \nonumber  \\
z_i (t+1) &=& z_i (t) ~~~~~~~~~~~~{\rm otherwise.}
\label{soam}
\end{eqnarray}
To avoid ambiguities, the random forces $\eta _{i,z}$ are now chosen
as real numbers in the interval $0 \le \eta < 1$. A very similar
model has been independently introduced by Roux and Hansen \cite{RouHan}.

We determine the roughness exponent $\zeta$ of the SOAM by simulating
interfaces according to the rules (\ref{43}), (\ref{soam})
and measure the width during a sufficiently long time interval for different
system sizes. The results \cite{PhD}
$\zeta(D=1) = 1.24 \pm 0.01$,
$\zeta(D=2) = 0.75 \pm 0.01$, and
$\zeta(D=3) = 0.34 \pm 0.01$ are consistent with the results of the
automaton model.

\subsection{Comparison with analytical results}

We now compare the numerical results with the analytical exponents
which are obtained by an extrapolation of the results
(\ref{41}) of the $\epsilon$-expansion (see Table 1).
It should be emphasized that the equation of motion (\ref{8})
is not a useful description of an interface in a two-dimensional
medium ($D=1$) because higher gradient terms are relevant.
This is in agreement with the unphysical
result $\zeta (D=1) \simeq 1.25 > 1$.
Nonetheless, the considered interfaces are stable and self-affine
because Eq. (\ref{8}) and
the automaton models (\ref{43}), (\ref{44}), and (\ref{soam})
do not allow overhangs. We include the case $D=1$ in Table 1
in order to compare the analytical and numerical results for
Eq. (\ref{8}).

The simulations support the analytical results:
In $D=3$ ($\epsilon =1$) the analytical results deviate from the numerical
values by only a few percent. In agreement with the expectation of an
$\epsilon$-expansion in $D=4-\epsilon$
the deviations become larger with decreasing $D$.
The numerical results in Table 1 are consistent with the exact scaling
relation (\ref{theta}).
However, the results for the roughness exponents
are clearly inconsistent
with the suggestion \cite {NarFisint} that $\zeta = (4-D) /3 $ is exact.
Assuming that there are no higher-order corrections to $\zeta$
\cite {NarFisint},
the numerical results indicate that there are non-perturbative corrections
to $\zeta$, a possibility mentioned by Narayan and Fisher \cite {NarFisint}.

The exactness of the result $\zeta = (4-D) /3 $ would have meant that
an interface at equilibrium ($F=0, ~T>0$) and a driven interface at
threshold $F_c$ have the same roughness exponent although the underlying
physics is different. The simulations show that the driven interface
is rougher than the interface in equilibrium.

The Harris condition (\ref{harcon})
for a sharp depinning transition is fulfilled
as an inequality by the numerical values.

\section {Summary}

In this paper we investigated the behavior of an elastic interface,
which is driven by a homogeneous force $F$ through a medium
with short-range correlated random forces.
It was made plausible that this problem can be described by the
equation of motion (\ref{8}) for a planar interface.
It was found that $D_c = 4$ is the upper critical dimension.
For $D>D_c$ and weak disorder, the interface is always smooth
and there is no threshold for driven motion if the random force
varies sufficiently slowly in space.
However, we showed by considering a mean field theory that, a
nonvanishing threshold $F_c > 0$ can be produced by a random force
with a cusp singularity ($\Delta '(0^+)<0$) of its correlator
$\Delta (z)$.
Below $D_c$ the interface becomes rough above the
Larkin length $L_c$. The correlation length $\xi$ is a second
important length scale, which diverges when approaching
the critical point $F_c$. The results of the perturbation theory
diverge with a power of $\xi$.

A simple three-parameter RG
is not a way out of these divergences, because the second moment
$Q_2 = \Delta ''(0) $ of the correlator
and therefore the inverse mobility $\lambda $ diverge
under renormalization at $L=L_c$.
Extending the space of renormalizable parameters to all
moments $Q_{2n} = \Delta ^{(2n)} (0)$ of the correlator, i.e.
to the whole {\it function} $\Delta (z)$, we found that the above
divergences correspond to a cusp singularity of $\Delta (z) $ at
the origin. With the interpretation
$Q_{2n} = \Delta ^{(2n)} (0^+)$ a consistent RG scheme could be
constructed. The cusp singularity $\Delta '(0^+) < 0$
produces a finite threshold $F_c$, similar to the
mean field theory \cite{hand}.
Above the threshold, the RG is stopped at the correlation length $L=\xi$,
above which one crosses over to a EW regime, where the random forces
act independently on the moving interface. The behavior of the
interface at different length scales $L$ and driving forces $F$
can be summarized in a phase diagram (see Fig. 9).

The exponents in the critical regime $L_c < L < \xi$ were calculated
to first order in $\epsilon = 4-D$ (see Eq. (\ref{41})).
Although the result for the roughness exponent
$\zeta \simeq \epsilon / 3 $ coincides with that of the equilibrium
problem to first order in $\epsilon$,
our numerical simulations show that the roughness at the
threshold and zero temperature is larger than that in equilibrium
($F=0, T>0$). As expected, the velocity exponent $\theta$ decreases
when the dimension is decreased, i.e. for a rougher
interface the onset of the motion at $F_c$ is more abrupt.
Our results for the correlation length exponent $\nu$
fulfill a Harris-like criterion to observe a sharp transition
as an equality for the first-order results and as an inequality for
the numerical results.

\section*{ Acknowledgements}

The research is partially supported
by the Deutsche Forschungsgemeinschaft through
Sonderforschungsbereich 166 and 341.

\begin{appendix}

\section{Velocity}

In this Appendix the perturbative calculation
of the velocity $v$ of the interface is discussed.
We give two examples how typical integrals for
the corrections to the velocity can be calculated.

\subsection{First-order correction}

The velocity $v$ can be obtained by iterating Eq. (\ref{10}):
The $n-th$ order iteration for $\tilde{h}({\bf k},w)$ follows from (\ref{10})
by using the $(n-1)-th$ order iteration on its $rhs$. (The
zeroth order iteration of $\tilde{h}$ vanishes.)
After truncating this procedure at
the order which is needed, the resulting equation is averaged over
the disorder, which gives on the $lhs$
$ \langle \tilde h({\bf k}, \omega) \rangle =0 $, while the $rhs$
is a function of $v$.
For example, the lowest-order correction (Fig. 4) to $ v_0 = F/\lambda $
is obtained
from the second-order iteration
\begin{equation}
\label{18}
{F - \lambda v  \over \lambda v} \simeq
-{1 \over \lambda v}
\int _q
\int _{\bf p} ~{iq~\Delta _q~ \tilde \Delta _\parallel ({\bf p})
\over
\gamma {\bf p}^2 -  i\lambda vq } \equiv I_1 ,
\end{equation}
where we have included the Fourier transform
$ \tilde \Delta _\parallel({\bf p}) $
of the correlator along the interface (see Eq. (\ref{12})).

First, we make $I_1$ real, perform the angular integration
and take the correlator $\tilde \Delta _\parallel ({\bf p})$ into account by
truncating the $p$-integration at
$1/a_{\parallel}$:
\begin{equation}
\label{Bgeskor}
I_1  \simeq K_D
\int _q q^2~\Delta _q~
\int ^{1/a_{\parallel}} dp {p^{D-1}
  \over
\gamma^2 p^4 +  (\lambda vq)^2 },
\end{equation}
where $K_D^{-1} = 2^{D-1} \pi ^{D/2} \Gamma (D/2) $.
For $D>4$ this integral converges for all $v$ as long as
$a_{\parallel} $ is finite.
For $D <4$, however,
the limit $v \to 0$ causes a divergence at small $p$.
If we set $a_{\parallel} = 0$,
the integral (\ref{Bgeskor}) diverges
at large $p$ when $4-D=\epsilon \to 0$.

After introducing the new variable
$y=p/\sqrt{\lambda v q /\gamma}$,
we get
\begin{equation}
I_1  \simeq {K_D \over \gamma ^{D/2}} (\lambda v)^{-\epsilon/2}
\int _q  q^{D/2}~\Delta _q~
\int ^{\sqrt{\gamma /\lambda v q}/a_{\parallel}} dy {y^{D-1}
  \over
y^4 +  1 } ~.
\end{equation}
The divergence with $v\to 0$ is now outside the integral and we can set the
upper limit to infinity and thereby getting the desired leading term
proportional to $1/\epsilon$:
\begin{equation}
I_1  \simeq -{K_D \over \gamma ^2} \xi_0^{\epsilon}
{}~\Delta ''(0)
\int ^{\infty}_1 dy~~ y^{D-5},
\end{equation}
where Eq. (\ref{13}) was used. Hence one gets the
result
\begin{equation}
\label{Cdefdl1}
{F - \lambda v  \over \lambda v} \simeq
cQ_2 {1\over \epsilon } \xi_0^{\epsilon}
\equiv \delta \lambda ^{(1)}/\lambda ,
\end{equation}
which is the same as Eq. (\ref{17b}).

However, for very large $ v~$, $\xi_0 \ll a_\parallel $,
i.e.
$ \sqrt{\lambda vq/\gamma} $ can no longer be regarded
as a lower cutoff for
the $ p $-integration. The latter has to be truncated at an
upper cutoff $ 1/a_\parallel $.
In this case one can introduce the new
variable $q'= \lambda v q / \gamma p^2 $. Then the integrals
in Eq. (\ref{Bgeskor}) can be carried out:
\begin{equation}
{F - \lambda v  \over \lambda v} \simeq
{K_D \over \lambda ^2 v^2} {a_\parallel ^{-D} \over D}
\int _{q'} \Delta _{q'} { q^{\prime 2} \over 1 + q^{\prime 2} }
\simeq \Delta (0) {K_D \over \lambda ^2 v^2} {a_\parallel ^{-D} \over D}.
\end{equation}
As expected, the correction to $ F - \lambda v $ vanishes for large $v$,
where Eq. (\ref{8}) is of EW-type.

\subsection{Second-order correction}

Next we extend the calculation of the velocity to second order.
To this end Eq. (\ref{10}) is iterated as described above.
The second order diagrams for the velocity are shown
in Fig. 14. Since we want to use the perturbation theory to
construct a renormalization group expansion in $ D= 4-\epsilon $
we only calculate the leading order terms in $ \epsilon $.
The expression for diagram 14a) is given by
$$ I_2 \equiv \int _q \int _{q'}
\int _{\bf  p } \int _{\bf p' }
{}~\Delta _q~\Delta _{q'}~
\tilde \Delta _\parallel ({\bf p})
\tilde \Delta _\parallel ({\bf p'})~
i^3~q^2 q' $$
\begin{equation}
\label{C7neu}
\times { 1 \over \gamma  {\bf p}^2 +  i\lambda vq } ~
 { 1 \over \gamma  {\bf p'}^2 +  i\lambda vq' }
{ 1 \over \gamma  ({\bf p+p'})^2 +  i\lambda v (q+q')} .
\end{equation}
The integrals will be evaluated below. The result is
\begin{equation}
\label{25}
\lambda v ~{1 \over 2}~\left ({\delta \lambda ^{(1)} \over \lambda}
\right ) ~.
\end{equation}
The calculation of diagram 14b) is very similar and
gives the same contribution as (\ref{25}).
Diagrams 14c) and 14d) correspond to a renormalization of the response function
and also contribute the term (\ref{25}).
The sum of the diagrams 14e)-g) is equal to zero, diagram 14h) cancels with
14j) and diagram 14i) cancels with 14k).
If we collect all diagrams up to second order we finally obtain
\begin{equation}
\lambda _ {eff} = \lambda \left[1 +
{\delta \lambda ^{(1)} \over \lambda}+
2\left({\delta \lambda ^{(1)} \over \lambda}\right)^2+...\right].
\end{equation}

\bigskip
As an example for the evaluation of second-order diagrams we calculate
the contribution (\ref{C7neu}) of diagram 14a).
The integrals over $q$ and $q'$ do not vanish
only if the powers of $q$ and $q'$ are even. Thus $I_2$ is
$$ I_2 = \int _q  \int _{q'}
\int _{\bf p } \int _{\bf p'}
{}~\Delta _q~\Delta _{q'}
\tilde \Delta _\parallel ({\bf p})
\tilde \Delta _\parallel ({\bf p'})~
q^2~q'^2 ~v$$
\begin{equation}
\label{A7}
\times { {\bf p}^2 {\bf p'}^2 + {\bf p}^2 ({\bf p+p'})^2
\over [\gamma ^2 {\bf p}^4 +  (\lambda vq)^2 ] ~
 [\gamma ^2 {\bf p'}^4 +  (\lambda vq')^2]}
{1 \over [ \gamma  ^2 ({\bf p+p'})^4
+  (\lambda v (q+q'))^2]}
\end{equation}
\begin{equation}
\label{BdefI_k}
\equiv \int _q \int _{q'}
{}~\Delta _q~\Delta _{q'}~q^2~q'^2 ~v ~~I_k.
\end{equation}
We consider here only the first term with ${\bf p}^2 {\bf p'}^2$
in Eq. (\ref{A7})
because the second one cancels with
a term of the diagram in Fig. 14b).
We have seen that the integration over $p$ gives the important
leading divergences, while the $q$-integration does not yield
divergent factors. Thus, for simplicity we set in
Eq. (\ref{BdefI_k}) $q=q'$,
and introduce new variables
${\bf y}={\bf p}/\sqrt{\lambda v q /\gamma}$, and
${\bf y'}={\bf p'}/\sqrt{\lambda v q /\gamma}$
and treat the limits
as in the calculation of the first order correction.
Then we obtain
\begin{equation}
I_k \simeq
\int ^{\sqrt{\gamma /\lambda v q}/a_{\parallel}}_1
{d^D y \over (2 \pi)^D }
\int ^{\sqrt{\gamma /\lambda v q}/a_{\parallel}}_1
{d^D y' \over (2 \pi)^D }~
{ 1 \over  {\bf y}^2 } ~
 { 1 \over {\bf y'}^2  }
 { 1 \over ({\bf y+y'})^4 },
\end{equation}
where the range of integration refers to the absolute values of
${\bf y} $ and ${\bf y'} $.
Again, the important divergence is at large $y$ and $y'$, so that
we can introduce a smooth cutoff for small $y$ and $y'$
and first perform the $y'$-integration
and set the upper limit to infinity:
$$ I_k \simeq
\int
_{\bf y }
{ 1 \over  ({\bf y}^2 +1)} ~ I'_k $$
with
\begin{equation}
I'_k=\int
_{\bf y'  }
 { 1 \over ({\bf y'}^2 +1) }
 { 1 \over [({\bf y+y'})^2 +1]^2}.
\end{equation}
This $D$-dimensional integral can be evaluated by introducing
a Feynman-parameter $x$ \cite {AMIT}
\begin{equation}
I'_k=\int
_{\bf y'} \Gamma (3)
\int _0 ^1 dx ~x
\left \{
 (1-x)({\bf y'}^2 +1) +
 x [({\bf y+y'})^2 +1] \right \} ^{-3}.
\end{equation}
Using the general formula
\begin{equation}
\int _{\bf y }
[ {\bf y}^2 + 2 {\bf y y'} + m^2]^{-\alpha}
= { K_D \over 2}
{ \Gamma \left ( {D \over 2} \right )
\Gamma \left ( \alpha - {D \over 2} \right ) \over
\Gamma (\alpha) }
\left [ m^2 - {\bf y'}^2 \right ] ^ {(D/2) - \alpha} ,
\end{equation}
one obtains
\begin{equation}
I'_k=
{ K_D \over 2}
 \Gamma (D/2)
\Gamma ( 3 - D /2  )
\int _0 ^1 dx ~x
\left \{ {\bf y}^2 (x-x^2) +1 \right \}^ {(D/2) - 3}.
\end{equation}
The $y$-integration is also performed by introducing a Feynman-parameter $x'$.
$$I_k \simeq
\left ({ K_D \over 2} \right )^2
 \Gamma ^2 ( D / 2 )
\Gamma ( \epsilon ) $$
\begin{equation}
\label{A13}
\times \int _0 ^1 dx
\int _0 ^1 dx'   ~x ~x^{\prime \epsilon /2}
\left \{ x'[x(1-x)-1]+1
\right \}^ {(\epsilon/2) - 2}.
\end{equation}
With $\epsilon /2 = \epsilon '\to 0$
this integral is logarithmically divergent.
The prefactor of this logarithmic divergence is the same as the
prefactor of the $1/\epsilon$ term for small $\epsilon$. Thus, we
consider only the integral Eq. (\ref{A13}) for
$\epsilon '= \epsilon/2  = 0$ and
calculate the desired prefactor. The integrals are then elementary
and the prefactor is calculated to one. Then we obtain with
$\Gamma (2 \epsilon ') \simeq 1/(2 \epsilon ')$
\begin{equation}
 I_k \simeq K_D ^2 {1 \over 2} {1 \over \epsilon ^2}.
\end{equation}
With Eq. (\ref{BdefI_k}) the result (\ref{25}) follows.

\section{Vertex correction}

In this Appendix we outline the calculation which leads to the
renormalization of moments resulting in Eq. (\ref{28}). The diagrammatic
representation of the bare moments $Q_{2n}$ is shown in Fig. 7a) , where all
incoming and outgoing momenta conjugated to ${\bf x}$ are set to zero.
According to Eq. (\ref{verana}) the analytical expression associated with the
diagram a) is exactly $Q_{2n}$. The diagrams contributing to the
renormalization of the moments up to the one-loop order are shown in Fig.
7b)-d). In each of these diagrams, one has only one integration over
the internal momentum conjugated to the space argument of the interface
height $h({\bf x})$. This integration has the form $\int_{\bf{p}}G_0^2(%
{\bf p})$ and yields for all diagrams under consideration the same factor
$c\xi _0^\epsilon /\epsilon $. Additionally to this integration there is
also an integration over the momenta $q_1$ and $q_2$ conjugated to the
argument of the disorder correlator $\Delta (z)$. The momenta $q_1$ and $q_2$
are associated with the dotted lines in one-loop diagrams in Fig. 7.

Let us consider the diagram b). The number of arrows starting at crossing
points of dotted lines with the continuous line is $n$ for both upper and
the down continuous lines. The distribution of arrows between the both
crossing point along a continuous line is arbitrary.
The sum of the arrows should obey
the condition $n_1 + n_2 =n $. According
to Eq. (\ref{10}) the factor $(\pm iq)^m$ is associated with each crossing
point with $m$ being the number of lines starting at the point under
consideration. The sign $+$ has to be taken if the dotted line belonging to
the crossing point goes out from the point, $-$ if the dotted line goes into
the crossing point. In the following we designate $-q$ by $\widetilde{q}$.
Summing over all possible distributions of arrows between the two
crossing points of the top line in the diagram b) we get the factor
\begin{equation}
i^{n+1}q_1(q_1+\widetilde{q}_2)^n.  \label{b1}
\end{equation}
The factor associated with the distribution of arrows on the down line in b)
is
\begin{equation}
i^{n+1}\widetilde{q}_1(\widetilde{q}_1+q_2)^n,  \label{b2}
\end{equation}
so that the factor associated with the whole diagram b) is
\begin{equation}
(-1)^{n+1}q_1\widetilde{q}_1(q_1+\widetilde{q}_2)^n(\widetilde{q}_1+q_2)^n.
\label{b3}
\end{equation}
Analogously the factor associated with the diagram c) is given by
\begin{equation}
(-1)^{n+1}q_1q_2(q_1+\widetilde{q}_2)^n(\widetilde{q}_1+q_2)^n. \label{b4}
\end{equation}
The factor associated with b) and c) is the sum of (\ref{b3}) and (\ref{b4})
\begin{equation}
(q_1+\widetilde{q}_2)^{2n+1}q_1=\sum_{k=1}^{2n+1}C_k^{2n+1}q_1^{2n+2-k}%
\widetilde{q}_2^k+q_1^{2n+2}q_2^0,  \label{b5}
\end{equation}
where $C_i^j={\bigl (}{_i^j}\bigr ) $ are the binomial
coefficients. The diagram d) in Fig. 7 gives the factor
\begin{equation}
-q_1^{2n+2}q_2^0,  \label{b6}
\end{equation}
which compensates the last term in Eq. (\ref{b5}). Thus, the part of the
diagrams b) - d) depending on $q_1$ and $q_2$ is given by
\begin{equation}
\sum_{k=1}^{2n+1}C_k^{2n+1}q_1^{2n+2-k}\widetilde{q}_2^k.  \label{b7}
\end{equation}
To obtain the analytical expression
associated with b)-d) we have to multiply
(\ref{b7}) with $\Delta (q_1)\Delta (q_2)$
and integrate over $q_1$ and $q_2$.
Taking into account that only the even moments of the disorder correlator
$\Delta (z)$ are non zero we get from (\ref{b7})
\begin{equation}
{\sum_{k=1}^nC_{2k}^{2n+1}Q_{2k}Q_{2n-2k+2}}.  \label{b8}
\end{equation}
The whole correction of b)-d) is obtained from (\ref{b8}) by multiplying the
latter with the factor $c\xi _0^\epsilon /\epsilon $. Finally, the
contribution from the diagrams a)-d) in Fig. 7 is
\begin{equation}
Q_{2n}+{\sum_{k=1}^nC_{2k}^{2n+1}Q_{2k}Q_{2n-2k+2}}c\xi _0^\epsilon
/\epsilon .   \label{b9}
\end{equation}
The latter gives the renormalization of the moments of the disorder
correlator up to the one-loop order.

\section{Functional flow equation}

In this Appendix it is shown how the vertex corrections of the moments
$Q_{2n}$
in Eq. (\ref{28}) can be put together in a functional flow equation
for $\Delta (z)$.

The flow equation can be formally obtained by differentiating Eq. (\ref{13})
\begin{equation}
\label{C1}
{\partial ~\Delta (z) \over \partial ~\ln L } =
 \sum _ {n=0} ^{\infty} {(iz)^{2n}\over (2n)!}
L {\partial Q_{2n}\over \partial L}~,
\end{equation}
where $Q_{2n}$ is identified with the effective moments
$Q_{2n,eff}$ which depend on the upper cutoff $L=\xi_0$.
Differentiating Eq. (\ref{28}) and writing $Q_{2n,eff} = Q_{2n}$ one gets
\begin{equation}
L {\partial Q_{2n}\over \partial L} =
c L^{\epsilon}
\sum _{j=1} ^n C^{2n+1}_{2j} Q_{2j} Q_{2(n-j+1)}.
\end{equation}
To construct a functional flow equation the $rhs$ of Eq. (\ref{C1})
has to be expressed as a function of $\Delta (z)$.
To this end one first extends the summation over $j$ to $0 \le j \le n$ and
subtracts the term with $j=0$ separately. Introducing the new
index $k=n-j$ and using
$C_{2j}^{2n+1} = C_{2(n-j)+1}^{2n+1}$,
the $rhs$ of Eq. (\ref{C1}) can be written as
\begin{equation}
\label{C3}
\sum _ {n=0} ^{\infty} {(iz)^{2n}\over (2n)!}
\sum _{j=0} ^n C^{2n+1}_{2j+1} Q_{2j+2} Q_{2n-2j}
- Q_0 \sum _ {n=0} ^{\infty} {(iz)^{2n}\over (2n)!} Q_{2n+2},
\end{equation}
where $j$ is written for $k$ and
the factor $cL^{\epsilon}$ is skipped.
With Eq. (\ref{13}) the second term can be expressed as
\begin{equation}
- {\partial ^2 \over \partial (iz) ^2 } \Delta (z) \Delta (0).
\end{equation}
The first term in (\ref{C3}) can be written as
\begin{equation}
{\partial \over \partial (iz)}  \left \{
{\partial \over \partial (iz') }\left .
\sum _ {n=0} ^{\infty} \sum _{j=0} ^n
{(iz')^{2j+2}\over (2j+2)!} Q_{2j+2}
\right |_{z'=z}
\label{C5}
{(iz)^{2n-2j}\over (2n-2j)!} Q_{2n-2j} \right \}.
\end{equation}
Writing down the sums over $n$ and $j$ term by term, one easily sees
that (\ref{C5}) is equal to
\begin{equation}
\label{C6}
{\partial \over \partial (iz)} \left \{
{\partial \over \partial (iz') }\left .
\sum _ {j=1} ^{\infty} {(iz')^{2j}\over (2j)!} Q_{2j}
\right | _{z'=z}
\sum _ {n=0} ^{\infty} {(iz)^{2n}\over (2n)!} Q_{2n} \right \} .
\end{equation}
The range of the summation over $j$ can be extended to
the vanishing term $j=0$.
Using Eq. (\ref{13}), (\ref{C6}) can be identified with
\begin{equation}
{\partial \over \partial (iz)} \left \{
\left . {\partial \over \partial (iz') }
\Delta (z')
\right | _{z'=z}
\Delta (z) \right \} =
- {1 \over 2}
{\partial ^2 \over \partial z ^2 } \Delta ^2 (z) .
\end{equation}
Now we can put the first and second term of Eq. (\ref{C3}) together
and insert it in Eq. (\ref{C1}):
\begin{equation}
{\partial ~\Delta (z) \over \partial ~\ln L } = - c L^{\epsilon}
{d^2 \over d z^2 } \left [ {1 \over 2} \Delta^2 (z)
- \Delta (z) \Delta (0) \right ],
\end{equation}
which is the functional flow equation (\ref{30d}).

\end{appendix}

\newpage


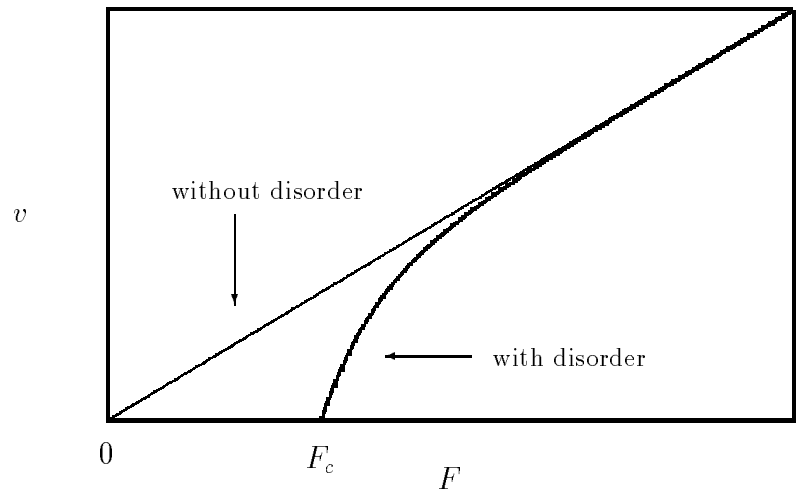
\begin{figure} \caption {Interface velocity $v$ as a function of
the driving force $F$.
}\end{figure}
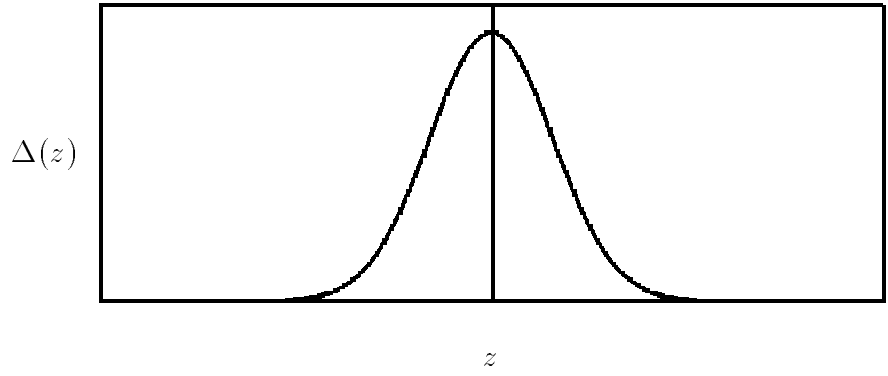
\begin{figure} \caption {The random force correlator $\Delta (z)$
for (a) random-fields, (b) random bonds, and (c) charge-density waves.
}\end{figure}
\begin{figure} \caption {
The diagrammatic representation of Eq. (22) (see text).
}\end{figure}
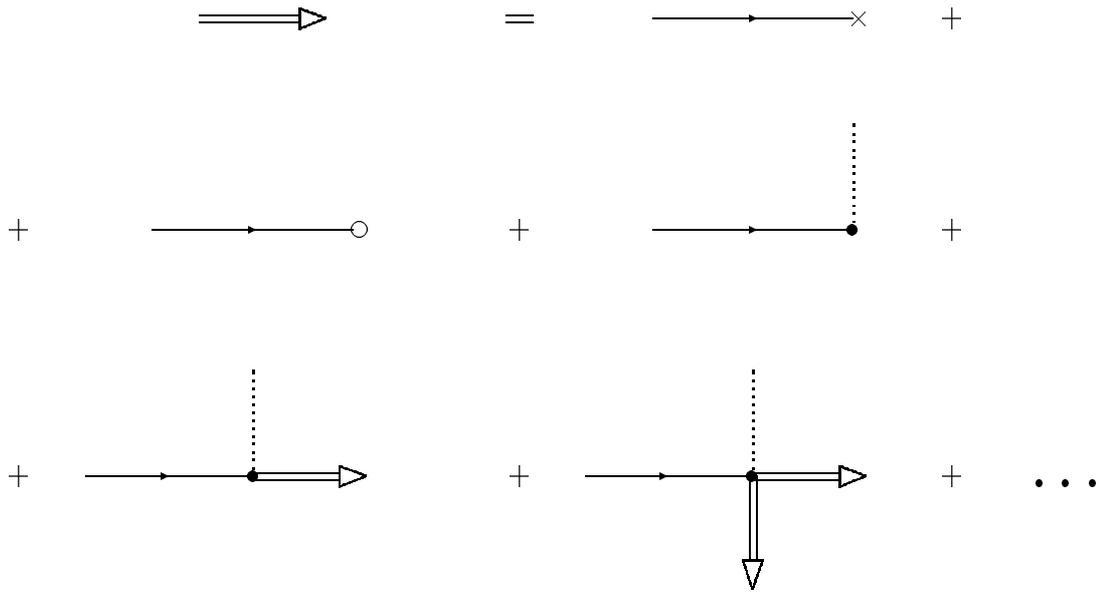
\begin{figure} \caption {
First-order correction to the velocity (see Eq. (29)).
The average over the quenched randomness is represented by the
connection of two outgoing dotted lines, which build up a ``loop''.
}\end{figure}
\begin{figure} \caption {
The response function $G({\bf k},\omega)$,
represented as a double line with an arrow,
and its one-loop expansion (see Eq. (33)).
}\end{figure}
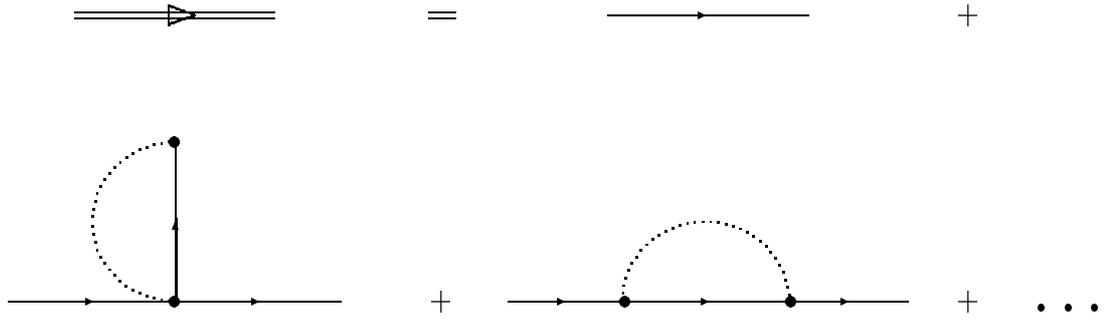
\begin{figure} \caption {
Lowest-order diagram
to the interface width (Eq. (39)).
}\end{figure}
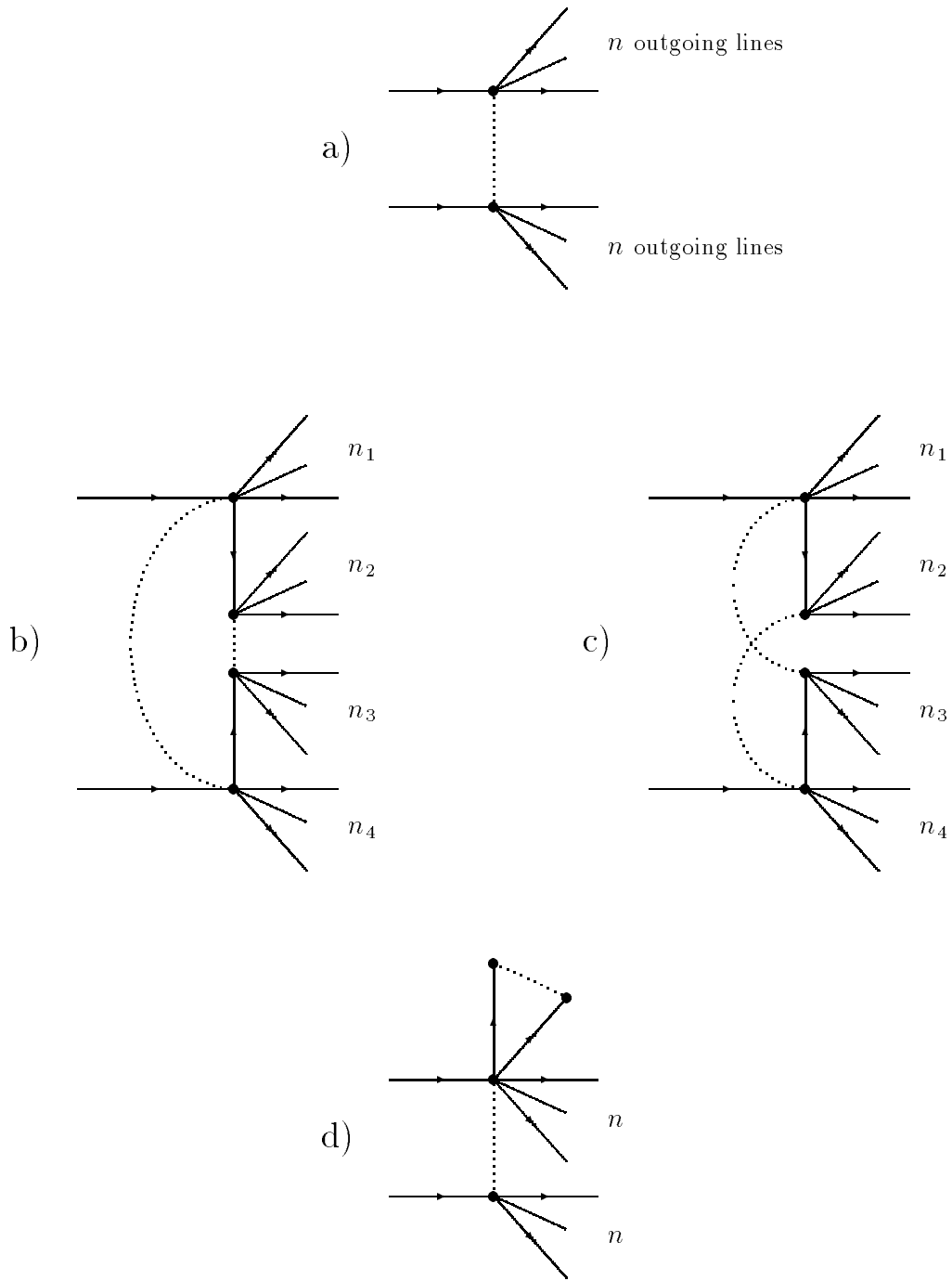
\begin{figure} \caption {
Diagrams illustrating the renormalization of the moments of $\Delta(z)$.
The diagrams shown contribute to the renormalization of $Q_{2n}$.
In diagram b) and c) the number of outgoing lines have to fulfill
the condition $n_1 + n_2 = n_3 + n_4 = n$.
}\end{figure}
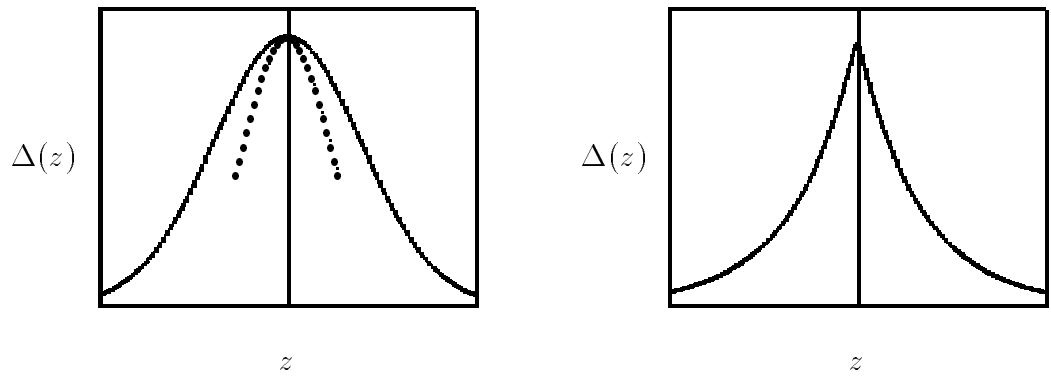
\begin{figure} \caption {
When starting with a smeared delta function (full line in figure on the $lhs$)
the second moment $Q_2 = \Delta ''(0)$ is negative. Under the
renormalization on scales $L<L_c$ the curvature
at the origin becomes greater and
$\Delta (z) $ becomes more peaked (dotted line). At $L\simeq L_c$ the
curvature diverges and the fixed-point solution has a cusp singularity
(figure on the $rhs$).
}\end{figure}
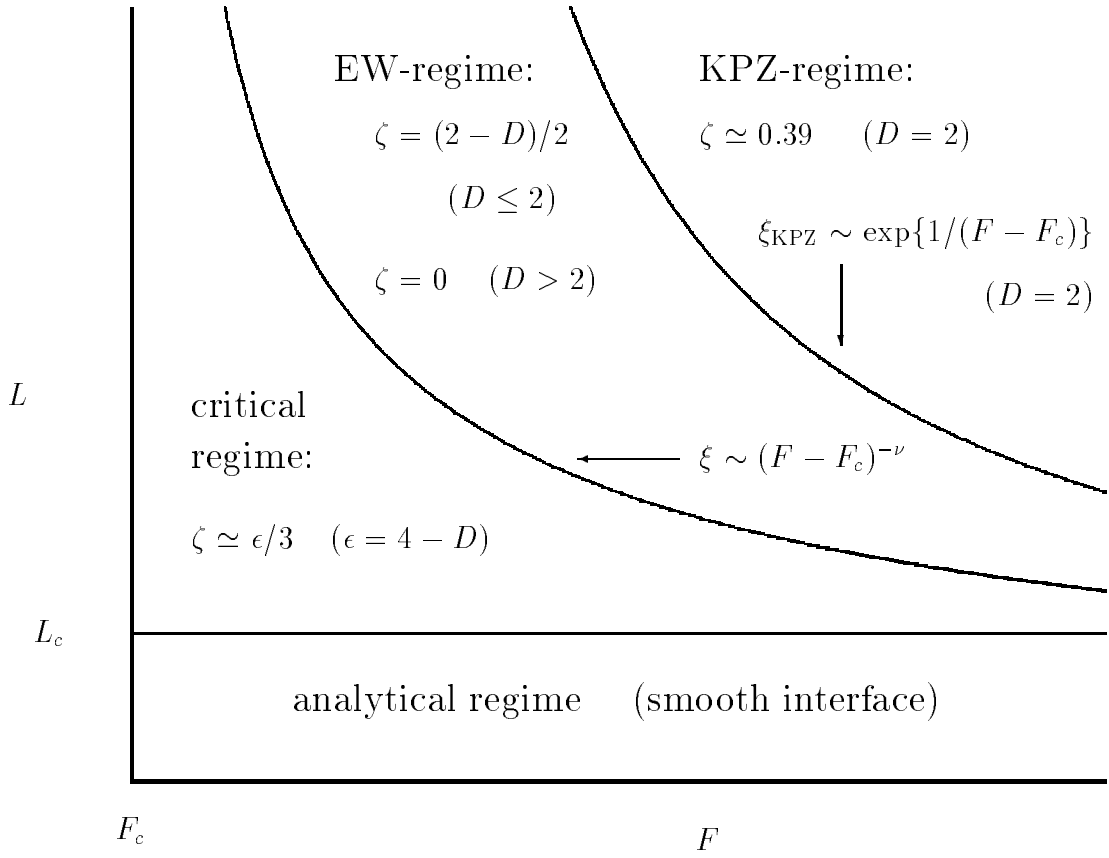
\begin{figure} \caption {$F-L$ phasedigram (see text).
}\end{figure}
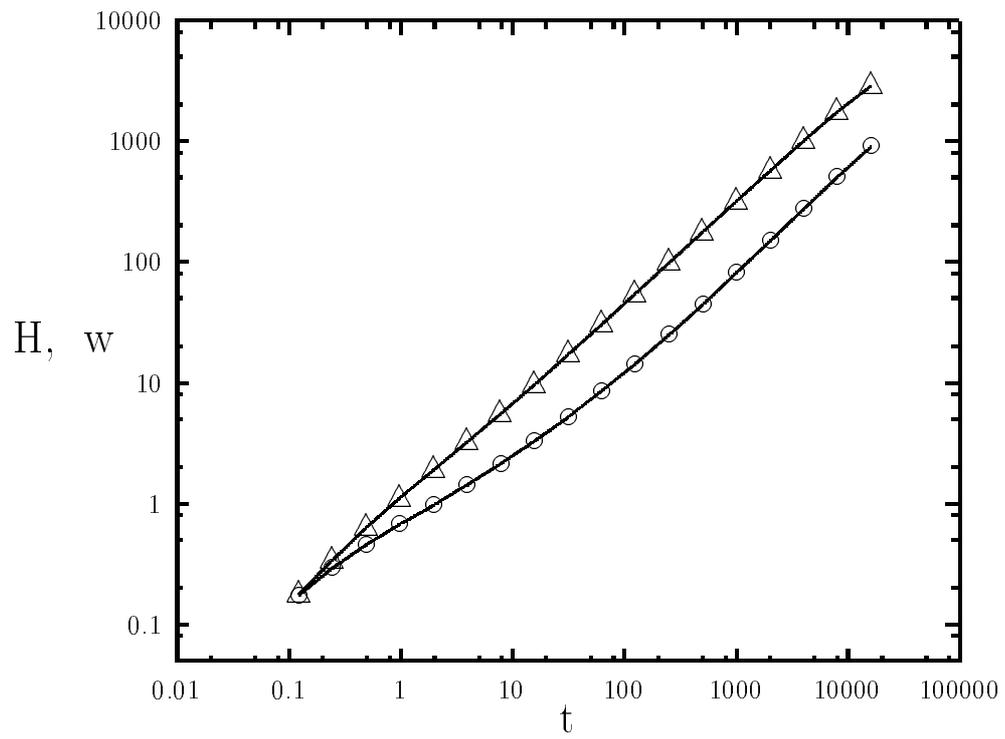
\begin{figure} \caption {
Scaling of the average position $H(t)$ (triangles) and interface width
$w(t)$ (circles) at threshold $F_c$
from a simulation of Eq. (62) in $D=1$. The system size is $L=16384$,
$\Delta t = 0.125$,  and the data
are averaged over 30 realizations of the disorder.
}\end{figure}
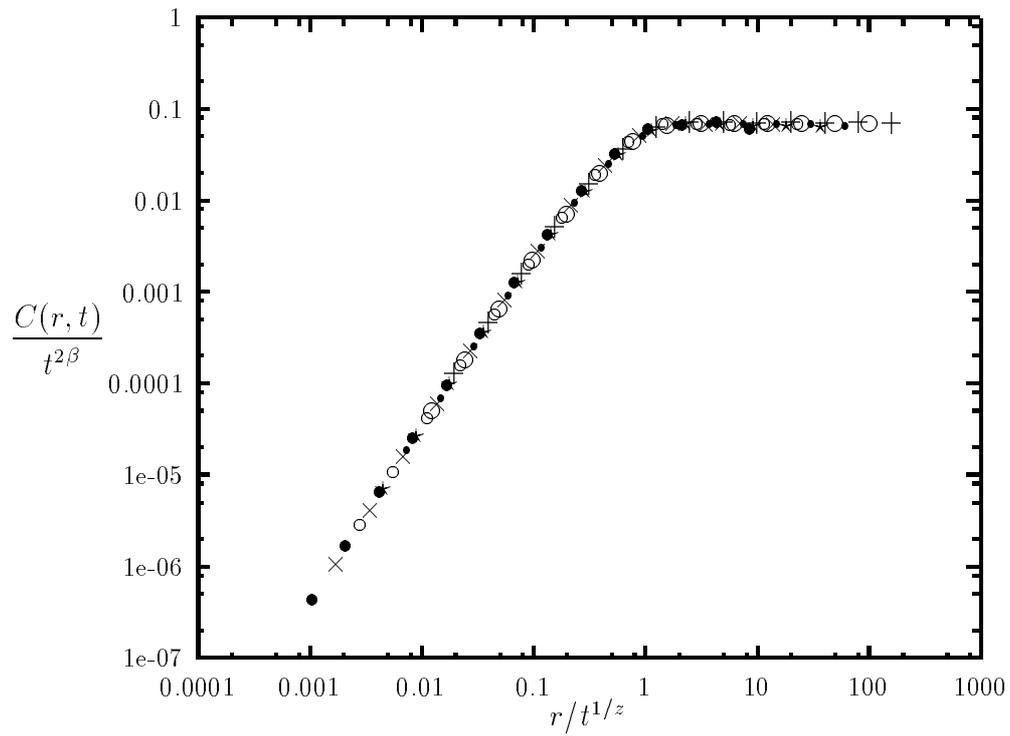
\begin{figure} \caption {
Scaling plot of the height-height correlation function
$C(r,t)$ from the same runs as in Fig. 10. The same plotting symbol is
used for data at a given time $t$.
}\end{figure}
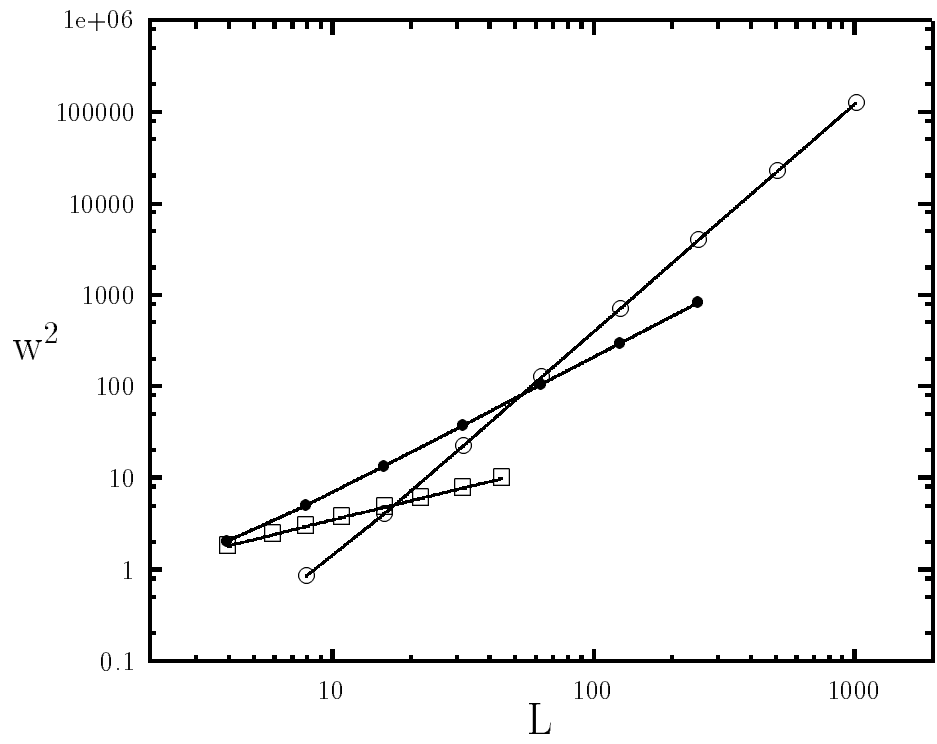
\begin{figure} \caption {
The width $w^2$ of pinned interfaces as a function of the system size $L$
from simulations of the automaton model for
$D=1,~g=1,~p=p_c \simeq 0.8004$ (open circles),
$D=2,~g=6,~p=p_c \simeq 0.74446$ (filled circles),
and $D=3,~g=8,~p=p_c \simeq 0.63165$ (squares).
}\end{figure}
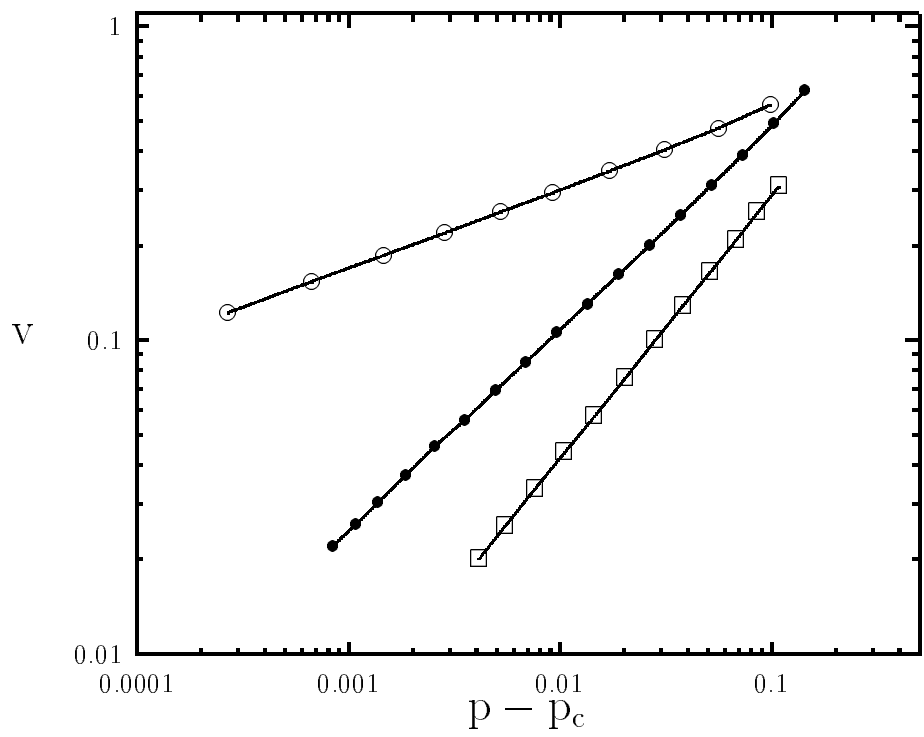
\begin{figure} \caption {
Interface velocity $v$ as a function of the driving force $p-p_c$
for $D=1$, $g=1$, $L \le 262144$ (open circles),
$D=2$, $g=6$, $L^2 = 1024^2$ (filled circles),
and $D=3$, $g=8$, $L \le 110^3$ (squares).
}\end{figure}
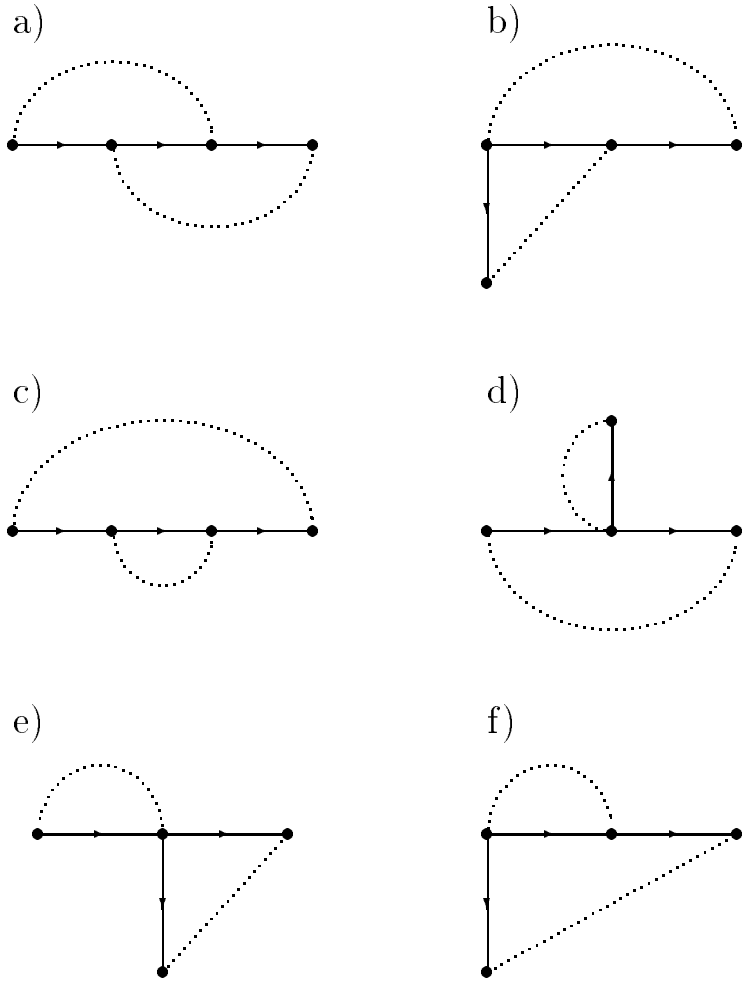
\begin{figure} \caption {
Second-order corrections to the velocity.
}\end{figure}

\begin{table}
\begin{center}
\begin{tabular}{|c|c||c|c|}\hline
\bf Exponent &\it D &\bf analytical &\bf numerical \\ \hline
& 1 & 1/3 & 0.25 $\pm$ 0.03  \\
velocity $\theta$ & 2 & 2/3 & 0.64 $\pm$ 0.02  \\
& 3 & 0.866.. & 0.84 $\pm$ 0.02  \\ \hline
& 1 & 1  & 1.25 $\pm$ 0.01 \\
roughness $\zeta$ & 2 & 2/3 & 0.75 $\pm$ 0.02  \\
& 3 & 1/3 & 0.35 $\pm$ 0.01  \\ \hline
& 1 & 3/4 & 0.88 $\pm$ 0.02 \\
roughening $\beta$ & 2 & 0.4286.. & 0.48 $\pm$ 0.01  \\
& 3 & 0.1875 & 0.20 $\pm$ 0.01 \\ \hline
& 1 & 4/3 & 1.42 $\pm$ 0.04 \\
dynamical $z=\zeta /\beta $ & 2 & 1.555.. & 1.56 $\pm$ 0.06 \\
& 3 &1.777.. & 1.75 $\pm$ 0.15  \\ \hline
& 1 & 1 & 1.33 $\pm$ 0.02  \\
correlation length $\nu$ & 2 &3/4& 0.80 $\pm$ 0.01  \\
$\nu = 1/(2 -\zeta)$ & 3 & 3/5 & 0.606 $\pm$ 0.004  \\ \hline
\end{tabular}
\end{center}
\caption {
Analytical and numerical results for the critical exponents
for interface dimensions $D=1,2$ and 3. The analytical values
are obtained by an extrapolation of the results (61)
of the $\epsilon$-expansion.
}\end{table}

\end {document}
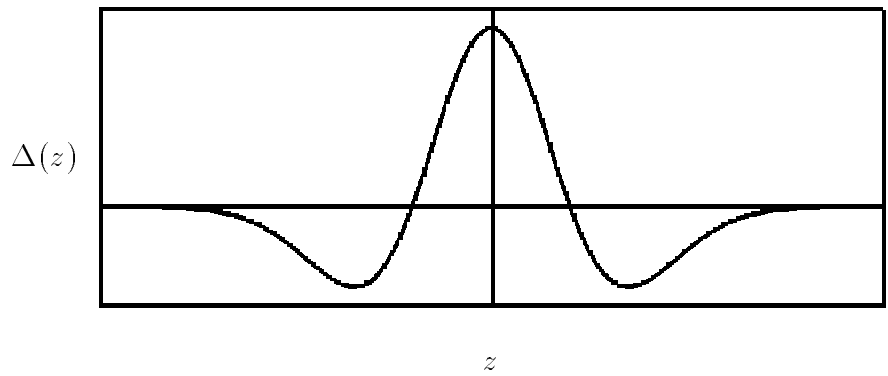
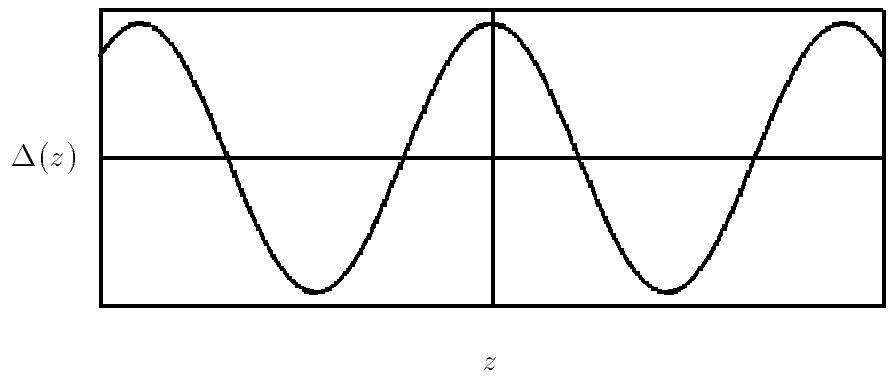
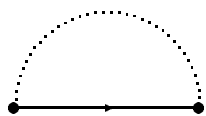
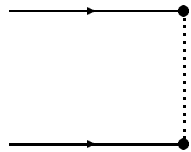
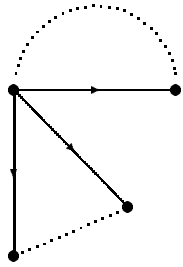
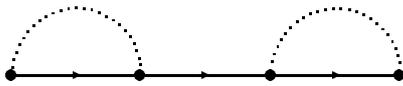
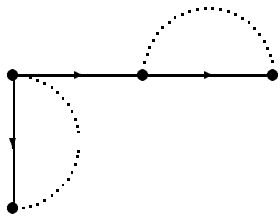
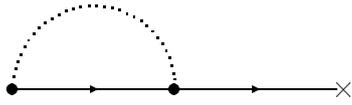
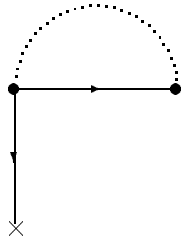
